\documentclass[12pt, twocolumn]{aastex62}

\received{December 10, 2019}
\accepted{February 16, 2010}
\submitjournal{AJ}

\begin{document}

\title{Cataclysmic Variables in the First Year of the Zwicky Transient Facility}

\correspondingauthor{Paula Szkody}
\email{szkody@astro.washington.edu}

\author[0000-0003-4373-7777]{Paula Szkody}
\affil{University of Washington,
Department of Astronomy, Box 351580,
Seattle, WA 98195, USA}

\author[0000-0002-8257-9727]{Brooke Dicenzo}
\affiliation{University of Washington,
Department of Astronomy, Box 351580,
Seattle, WA 98195, USA}

\author{Anna Y. Q. Ho}
\affiliation{Division of Physics, Mathematics and Astronomy,
California Institute of Technology,
Pasadena, CA 91125, USA}

\author{Lynne A. Hillenbrand}
\affiliation{Division of Physics, Mathematics and Astronomy,
California Institute of Technology,
Pasadena, CA 91125, USA}

\author[0000-0002-2626-2872]{Jan van Roestel}
\affiliation{Division of Physics, Mathematics and Astronomy,
California Institute of Technology,
Pasadena, CA 91125, USA}

\author{Margaret Ridder}
\affiliation{University of Washington,
Department of Astronomy, Box 351580,
Seattle, WA 98195, USA}

\author{Isabel DeJesus Lima}
\affiliation{University of Washington,
Department of Astronomy, Box 351580,
Seattle, WA 98195, USA}
\affiliation{Instituto Nacional de Pesquisas Espaciais
(INPE/MCTIC), Av. dos Astronautas, 1758 Sao Jose dos Campos, Brazil}

\author[0000-0002-9154-3136]{Melissa L. Graham}
\affiliation{DIRAC Institute, University of Washington,
Department of Astronomy, Box 351580,
Seattle, WA 98195, USA}

\author[0000-0001-8018-5348]{Eric C. Bellm}
\affiliation{University of Washington,
Department of Astronomy, Box 351580,
Seattle, WA 98195, USA}

\author{Kevin Burdge}
\affiliation{Division of Physics, Mathematics and Astronomy,
California Institute of Technology,
Pasadena, CA 91125, USA}

\author{Thomas Kupfer}
\affiliation{Kavli Institute for Theoretical Physics, University of California Santa Barbara, Santa Barbara, CA, USA}

\author[0000-0002-8850-3627]{Thomas A. Prince}
\affiliation{Division of Physics, mathematics and Astronomy, California Institute of Technology, Pasadena, CA 91125, USA}

\author{Frank J. Masci}
\affiliation{IPAC, California Institute of Technology, Pasadena, CA 91125, USA}

\author[0000-0001-7016-1692]{Przemyslaw J. Mr\'oz}
\affiliation{Division of Physics, Mathematics and
Astronomy, California Institute of Technology,
Pasadena, CA 91125, USA}

\author[0000-0001-8205-2506]{V. Zach Golkhou}
\affiliation{University of Washington \\
Department of Astronomy, University of Washington, Box 351580 \\
Seattle, WA 98195, USA}
\affiliation{The eScience Institute, University of Washington, Seattle, WA 98195, USA}

\author{Michael Coughlin}
\affiliation{LIGO Laboratory, California Institute of Technology, Pasadena, CA 91125, USA}

\author{Virginia A. Cunningham}
\affiliation{Department of Astronomy, University of Maryland,
College Park, MD 20742, USA}

\author{Richard Dekany}
\affiliation{Caltech Optical Observatories, California Institute of Technology, Pasadena, CA 91125, USA}

\author{Matthew J. Graham}
\affiliation{Division of Physics, Mathematics and Astronomy, California Institute of Technology,
Pasadena, CA 91125, USA}

\author{David Hale}
\affiliation{Caltech Optical Observatories, California Institute of Technology, Pasadena, CA 91125, USA}

\author{David Kaplan}
\affiliation{Department of Physics, University of Wisconsin-Milwaukee, Milwaukee, WI 53201, USA}

\author{Mansi M. Kasliwal}
\affiliation{Division of Physics, Mathematics and Astronomy, California Institute of Technology,
Pasadena, CA 91125, USA}

\author{Adam A. Miller}
\affiliation{Department of Physics and Astronomy, Northwestern University, Evanston, IL 60208, USA}

\author{James D. Neill}
\affiliation{Division of Physics, Mathematics and
Astronomy, California Institute of Technology,
Pasadena, CA 91125, USA}

\author[0000-0002-4753-3387]{Maria T. Patterson}
\affiliation{High Alpha, 55 Monumnet Circle Suite 1400, Indianapolis, IN 46204, USA}

\author{Reed Riddle}
\affiliation{Division of Physics, Mathematics and Astronomy, California Institute of Technology,
Pasadena, CA 91125, USA}

\author{Roger Smith}
\affiliation{Division of Physics, Mathematics and Astronomy, California Institute of Technology,
Pasadena, CA 91125, USA} 
 
\author[0000-0001-6753-1488]{Maayane T. Soumagnac}
\affiliation{Lawrence Berkeley National Lab, 1 Cyclotron Road, Berkeley, CA 94720, USA}
\affiliation{Department of Particle Physics and Astrophysics, Weizmann Institute of Science, Rehovot, Israel}

\begin{abstract}
Using selection criteria based on amplitude, time and color, we have identified
329 objects as known or candidate cataclysmic variable (CVs) during the first year of testing and operation of the Zwicky Transient Facility (ZTF). Of these, 90 are previously confirmed CVs, 218 are strong candidates based on the shape and color of their light curves obtained during 3-562 days of observations, and the remaining 21 are possible CVs but with too few data points to be listed as good candidates. Almost half the strong candidates are within 10 deg of the galactic plane, in contrast to most other large surveys which have avoided crowded fields. The available Gaia parallaxes
are consistent with sampling the low mass transfer CVs, as predicted by population models. Our followup spectra have
confirmed Balmer/helium emission lines in 27 objects, with four showing high excitation HeII emission, including candidates for an AM CVn, a polar and an intermediate polar. Our results demonstrate that a complete survey of the galactic plane is needed to accomplish an accurate determination of the number of CVs existing in the Milky Way.

\end{abstract}


\section{Introduction} 
The Zwicky Transient Facility (ZTF) is a northern
all-sky survey that uses the Palomar 48-inch telescope
equipped with a 47 deg$^{2}$ field
of view camera \citep{B19a,B19b,G19,M19}.
The advantages of the ZTF survey compared to past and on-going northern 
sky surveys are the combination of large sky coverage that includes the 
Galactic plane, along with color information
and increased temporal coverage, and the availability of nightly alerts. 
These aspects are ideal for finding variable stars, especially those that have
non-periodic, erractic brightness changes such as cataclysmic variables (CVs),
which consist of close binaries with mass transfer
from a late main-sequence secondary to a white dwarf
(see \citet{W95} for a review of all types of CVs). Combining the results
from all sky surveys along with astrometric data from Gaia will lead to the 
correct number and density of CVs throughout
our Galaxy, thus ultimately constraining population models and
close binary evolution scenarios. 

The ZTF project
uses 40\% of the time for the public, 40\% for partnerships, and 20\% for Caltech. As part of the
public portion, the available sky is sampled in ZTF $g$ or
$r$ filters every 3 nights and the available Galactic plane with 
$\mid b\mid\leq$7$^{\circ}$
every night in both $g$ and $r$ with mean 5$\sigma$ limiting magnitudes of 21 and saturation near 15 mag. Alerts on all variable 
objects are available every night. Commissioning took
place in the latter part of 2017 with the official
start of the 3 year survey on March 18, 2018 and the first public data release
DR1 occurred on May 8, 2019. DR1 is available from IPAC 
\footnote{https://www.ipac.caltech.edu/projet/ztf}.  

The alerts are constructed from difference images
between a reference field (consisting of a minimum
of 15 images) and the new field. These alerts
pass through the GROWTH Marshal \citep{K19} which
uses various filters constructed by participants to select specific types of
variables and transients. One such filter was
created to select candidate cataclysmic variables (CVs). In the CVs with a 
non-magnetic white dwarf (B$<$1 MG), the mass transfer accumulates in
a disk before accreting onto the white dwarf. The amount of mass transfer and 
accretion determines what the light curve of a specific CV will look like. 
Those with relatively low transfer rate values will undergo a
disk instability that results in a dwarf nova outburst (a rise in brightness of 2-9 mag within 1-3 days, the high brightness lasting for 1-15 days, and
a subsequent return to quiescence), with a repetition timescale
of weeks to months. The lowest mass transfer systems are those with the largest amplitudes, the longest times
at outburst and the longest
recurrence time between outbursts (years). The highest rates of mass transfer 
result in novalike (NL) systems, which have no outbursts but sometimes
undergo several magnitude transitions between low and high accretion states that can last for weeks at a time. In
systems with a magnetic white dwarf, the inner disk region can be channelled to the magnetic poles via
accretion curtains (called intermediate polars; IPs) or in the highest field 
cases (termed polars), the mass transfer goes directly
to the white dwarf magnetic poles, and no disk exists. 
Without a disk, there are no outbursts
but high and low states of accretion can occur, resulting in several magnitude 
changes in their light curves
on timescales of weeks. The colors of CVs are generally blue (ZTF $g-r$ close to 0 and
even bluer during an outburst) due to the contributions of the hot white dwarf 
and the accretion disk. However, long period systems with K type secondaries or those with
magnetic white dwarfs and resulting cyclotron emission from the accretion pole can be redder in color.

To find CVs, a simple GROWTH filter was used to search for 
non-moving point sources with an amplitude change of
2 mags within a timespan of 3 days and a color (from PanSTARRS)
of $g-r<0.6$. A real-bogus (rb) filter (calibrated from a zooniverse
program of human classification on a large dataset of ZTF images) was set to be
 low (0.1) to maximize findings. The filter began its full operation on June 5, 2018. With this filter, each night resulted in anywhere from
30-200 objects, the number mostly dependent on the weather and partly on
the location of the observed fields. The resulting objects were
then searched by eye to identify possible CVs. After many months of data, it 
became clear that it was most worthwhile to do the eye search if the rb was
greater than 0.5. Future refinements to the filter and machine learning
classifiers being developed by other teams in the ZTF project will
undoubtedly be able to decrease the amount of human interaction required. 
This paper presents the resulting previously confirmed CVs and CV candidates 
found using the simple GROWTH Marshal filter up to the time of the first
public data release.

\section{Identifying CVs} 

Each night, the combined $g,r$ ZTF light curves of the filtered objects provided by the Marshal were scanned by eye
to determine a possible dwarf nova
outburst or a change in state of a novalike system within the 30 day interval
depicted by the Marshal. Likely candidates were saved and checked against
known sources via SIMBAD, the Sloan
Digital Sky Survey \citep{Y00}, the Catalina Real-time
Transient Survey (CRTS;\citep{D09,D14}, MASTER \citep{L10} and
ASAS-SN \citep{Sh14}. The saved sources
continued to be monitored to help determine the correct classification. 
In many cases, this
allowed full outbursts to be observed, in other cases, only rises or declines from an outburst or high/low state were viewed. 
Other ZTF groups also transferred objects that passed their different filters (e.g. nuclear transients, SN, etc) if they thought they might be CVs. 
When the weather allowed long stretches of good observations, it was common to end with 50-70 previously known CVs and/or good  candidates per month. The most frequent contaminants were RR Lyrae and other
large amplitude periodic variables that fell within the color range.

Followup 
spectra on some objects were obtained using several telescopes, sometimes with multiple spectra on the same object (27 systems with the Palomar
60-in, 5 with the Palomar 200-in, 10 with the Keck 10m, 4 with the William Herschel 4.2m (WHT), 14 with the Apache Point Observatory (APO) 3.5m,
and 2 with the Liverpool 2m 
telescope) to confirm candidates. The
presence of Balmer emission (from the decline or quiescent state) or the presence of \ion{He}{2} emission (indicative
of a magnetic white dwarf or a very high mass transfer system) were used as confirmation criteria. Spectra obtained too close to an outburst generally 
showed only Balmer absorption lines and are indicative of an accretion disk
but were not used as confirmation.

The SED Machine \citep{Bl18,R19}, a low resolution (50\AA) integral field spectrograph operating from 3650-10,000\AA\ obtained spectra on the Palomar 60-in telescope. An automatic pipeline reduced the data and uploaded it to the Marshal. The Double Beam Spectrograph \citep{OG82} was used on the Palomar 200 in, the Low Resolution Imaging Spectrometer (LRIS; \citep{O95}) on the Keck and the Spectrograph for the Rapid Acquisition of Transients (SPRAT) on the
Liverpool telescope \citep{St04}, and uploaded after reduction pipelines. A few spectra were obtained at medium resolution (3.3\AA) with the WHT using the ACAM spectrograph \citep{B08}.
The APO data were obtained with the Double Imaging Spectrograph using either low resolution  (2\AA)
or high resolution (0.6\AA) gratings to cover the Balmer lines and the data reduced using $\it{IRAF}$ routines to
calibrate the data via He, Ne, Ar lamps
and flux standards obtained during each night. The APO blue CCD suffered contamination problems throughout the year, so most of the data collected were only from the red side of the dichroic (5500-9000\AA).

\section{Results}

The yearlong scans of the GROWTH Marshal with the CV filter
yielded 90 previously confirmed (from spectra or from the presence of a superhump outburst feature
\citep{W95}) CVs, 218 strong candidates based on their light curves and 21 objects which might be CVs but the data are too limited to tell for sure. Table 1 lists
the previously confirmed objects, Table 2 the
strong candidates and Table 3 the remaining possibilities. Some sources were listed as candidates in CRTS or MASTER, but if they have not been confirmed, we placed them in Table 2. In the rest of this paper, we
will generally refer to objects by their abbreviated
RA(HHMM) and Dec(Deg) i.e. ZTF0014+59, but provide RA(HHMMSS) if needed to differentiate sources. The tables also
provide the Galactic latitude, the range in magnitudes from outburst peak to quiescence, or from high to low accretion states observed by ZTF, the Gaia parallax and errors in mas (for those objects
with a measurement that was more than 3 times the error), the
number of normal outbursts and longer superoutbursts (SOBs) observed in the Marshal light curves, the
interval of actual coverage by ZTF in days up to 2019 May 8, if the source is in
the Sloan Digital Sky Survey (SDSS) footprint, the number of outbursts visible in CRTS if covered by
that survey, any spectra obtained with the ZTF instruments or available from the SDSS or the 
literature, and any
other relevant information.

The easiest systems to classify as CVs
are those that show a long string of observations at quiescence along with one SOB that has a large amplitude (typically 5-9 magnitudes) and a long duration (typically 3 weeks). A typical example (ZTF1727+23) is
shown in Figure 1. The SOB often has a distinct
shape, with a linear slowly declining plateau following the peak magnitude that lasts for 1-2 weeks, then a steep decrease toward the quiescent magnitude. In a few cases, there
are one or more rebrightening episodes during the decline (examples are ZTF2231+33 with one rebrightening and ZTF184159+37 with 6 possible rebrightnenings shown in Figure 1). These
systems are known as WZ Sge stars \citep{W95} or
Tremendous Outburst Amplitude Dwarf novae
(TOADS) \citep{HSC95}. Modeling of TOADS has shown that they are the systems with the lowest accretion rates among dwarf novae \citep{HSC95} and the ones that population synthesis models \citep{HNR01} predict should be the most numerous in surveys of CVs. They should have evolved to  the shortest orbital periods in their evolution during the
lifetime of the Galaxy. Most dwarf novae that have orbital periods less than
2 hrs show both SOBs and normal outbursts. The
systems with larger accretion rates are generally
those with longer orbital periods and with short recurrence times
for outbursts, so they are readily found in sky surveys with coverage of
weeks to months. An example of this kind of system is ZTF0613+06 in Figure 1.
All objects in Table 2 show outbursts with one of the types described above.
There also exist novalike systems which do not show outbursts but do exhibit high and low states
of accretion that last for many days. Two of these previously known systems, ZTF0854+39 (FR Lyn), and ZTF1631+69 (shown
in Figure 2) were found in the Marshal, but none of the candidates in Tables 2 or 3 showed this behavior.
The filter works best to detect outburst behavior and will need to be modified to pick up the generally slower
transitions between high and low states.

For the most part, the objects in Table 3 did not have enough data to unambiguously
determine if their variability is due to a dwarf nova outburst or to some type
of pulsational instability. As the survey proceeds, more observations may provide
an unambiguous classification.

\subsection{Spectroscopic Confirmations}
Many of the SEDM spectra were obtained near outburst, when the accretion disk is dominant and producing a
blue continuum or weak absorption lines. Thus, the low resolution usually only showed the continuum. The higher
resolution and larger telescopes of the Palomar 200 in, Keck, WHT and APO enabled better determination of Balmer and Helium
emission lines as the systems evolved to quiescence. In addition, 2 candidates were found to have spectra available from SDSS. In total, 29 of the systems in Table 2 were able to be positively confirmed as CVs with noticeable Balmer or Helium emission lines (those designated with a small letter e in the Spec column of Table
2).  Figure 3 shows the blue and red emission line spectra from the ZTF Marshal followup (coverage 
from 4000-9000\AA) while Figure 4 shows the red region from 5500-9000\AA) available from APO data when the blue side was not operational.

\subsection{The Galactic plane}
One of the unique features of ZTF is its
frequent observations within 7$^{\circ}$ of the Galactic plane (see
the spatial distribution of single-exposure epochs in DR1 
\footnote{https://www.ztf.caltech.edu/page/dr1}. Previous all-sky surveys such
as CRTS \citep{D14} generally avoided latitudes
within 10$^{\circ}$ of the galactic plane due to crowding and
large pixel sizes. Exceptions include a few targeted surveys such as the Optical Gravitational
Lensing Experiment (OGLE) that were successful in finding large numbers of dwarf
novae in the fields toward the Galactic Bulge and the Magellanic Clouds \citep{M15}
and the All Sky Automated Survey (ASAS) which had large pixels and a faint limit of only 14 mag \citep{P97}.
The 1 arcsec pixels of ZTF allow the
nightly plane observations to provide variable sources that can be identified in
the Marshal. Figure 5 shows the distribution of galactic
latitudes for the objects in Tables 1 and 2. The number of objects lying within 10$^{\circ}$ of the galactic plane
confirm the higher densities expected there, albeit not as large as the numbers
 found from OGLE (likely due to the limitations expressed in Section 5). Of the previously identified CVs (Table 1), only
23 (26\%) are within 10$^{\circ}$, while among the strong candidates (Table 2), the
number is 98 (45\%) and 7 (33\%) for the possible candidates in Table 3.
While the more frequent sampling of the plane by ZTF maps the
shape of outbursts better than the 3 day sampling, the detection rates of
outbursts should be about the same over the sky as dwarf nova outbursts 
normally last longer than 3 days. Thus, the summed number of known plus 
candidate systems (30-40 per 10$^{\circ}$ bin in the plane) compared to 
$\leq$ 20 per bin 
out of the plane does imply a higher density of CVs, with most being new 
candidates.

\subsection{Absolute Magnitudes}

The available Gaia parallaxes \citep{G18} provide distances that enable 
meaningful absolute magnitudes and 
heights above the plane without relying on average
absolute magnitudes at quiescence or outburst. However, the sample is small
due to the current limits on the Gaia parallaxes. Table 1 (confirmed CVs) 
has 57 (63\%)
with good (S/N$>$3) parallax measurements, while Table 2 has only 29 (13\%) and Table 3 has 7 (33\%).
These numbers are consistent with the closest systems being found in 
previous discoveries
and the new candidates having large distances and/or fainter brightness. 
 Figure 6 plots the absolute magnitude of
the objects from Tables 1 and 2. In the cases where
ZTF did not get a detection for the quiescent magnitude, the upper limits mean that the absolute magnitudes will
be even fainter than shown. Past results on absolute magnitudes of dwarf novae at quiescence \citep{W87,W95} have
shown a range from 7.5-11, depending on the orbital period and the outburst
recurrence time. Figure 6 shows that
 the majority of the ZTF sources are between 10-12, confirming that most systems
are faint. To investigate whether ZTF is predisposed to a particular
outburst type that would be related to mass transfer rate (such as the
low mass-transfer rate TOADS with infrequent, large amplitude outbursts
 \citep{HSC95}, the absolute magnitudes are plotted versus detected outburst
frequency in the interval covered by ZTF, in Figure 7 (top), as well as  
versus outburst amplitude (Figure 7 bottom). While the majority of 
the objects have outburst 
frequencies less than 0.02 (50 days), there is a large scatter at all
brightnesses. This result is consistent with those from smaller datasets (e.g. \citep{HSC95,T02}
that showed a wide range of outburst behavior for similar orbital 
periods and luminosities. However, the outburst amplitudes show a clear 
increase as the
absolute magnitudes become fainter. Of the 14 objects fainter than 12th
mag, 8 would qualify as candidates for TOADS \citep{HSC95}, with outburst
amplitudes $\geq$6 mag, and modeled by \citet{HSC95} with low
mass transfer and low disk viscosity. These 8 are all in the previously
known CVs (Table 1), while candidate CVs all have brighter absolute
magnitudes. This is likely related to the low frequency of outbursts
evident in this faint magnitude range (Figure 7 top) and the short 
timescale of the ZTF survey.

\section{Notes on Individual Systems}
Brief descriptions of the systems
with interesting features in their spectra or lightcurves are provided below.

\subsubsection{\ion{He}{2} objects}

Figure 3 shows four systems from Table 2 with strong \ion{He}{2}4686, or only helium lines, characteristics
of AM CVn systems or those containing a magnetic white dwarf and/or very
high accretion. ZTF2128+63 shows only lines of He and so is likely an
AM CVn system. ZTF2123+15 has \ion{He}{2} stronger than H$\beta$ and a prominent blue
continuum, making it a strong candidate for 
either an IP or a member of the class of high mass transfer rate novalikes called SW Sex stars \citep{T91,H98} that have orbital periods between 3-4 hrs.
ZTF0451+16 also shows strong Balmer and \ion{He}{1} lines as well as relatively strong \ion{He}{2}, but without the blue continuum that is characteristic of a high mass accretion rate system, so is a candidate for
a polar system. ZTF1647+43 is peculiar, as it has a very strong \ion{He}{2} line and a very
blue continuum but very weak Balmer emission. Further data are needed to refine its classification.

ZTF1631+69 (Table 1) has been identified as a CV \citep{A98}, and is reported in the ROSAT and XMM catalogs,
but there is no detailed study of this system available in the literature. The ZTF light curve
shows the existence of high and low states, a common feature of high accretion rate SW Sex
systems as well as those containing highly magnetic white dwarfs (polars and IPs). A series of 5 sequential spectra obtained over the span of an hour at APO on
May 24, 2019 shows strongly doubled  Balmer lines along with \ion{He}{2}, as well as large
changes from one spectrum to the next (Figure 8). The velocities of the H$\alpha$ and H$\beta$ lines show a portion of a sinusoidal variation over 200 km s$^{-1}$ during the hour but the timespan is too short to determine an orbital period. The doubled lines are a signature of an accretion disk while the high excitation could indicate a magnetic white dwarf. Further data are needed to ascertain if this is an IP system.

\subsubsection{Strong Balmer lines}
Figures 3 and 4 show that ZTF0422+29, ZTF0429+18, ZTF0439+18, 
ZTF0914+67, ZTF1315+42, ZTF1800+15, ZTF1704+26, ZTF1800+15, ZTF1829+31, ZTF1857+32, ZTF1900+30, and ZTF2248+38 all have the prominent Balmer emission lines typical of quiescent dwarf novae. ZTF0819+21, ZTF1037+12, ZTF1053+28, ZTF1716+29, ZTF1726+36, ZTF1810+32, ZTF1834+54, ZTF1841+37,
ZTF1954+18, ZTF2227+55 and ZTF2256+35 show weak H$\alpha$ in
emission while the bluer Balmer lines are in absorption (for those with blue spectra as well as the red). The spectra of these systems were obtained close to their outburst brightness and reflect the prominence of the thick accretion disk at those times.

\subsubsection{Peculiar Lightcurves}

ZTF1841+37 (Figure 1) shows a SOB followed by 6 normal outbursts or rebrightenings within 60 days. If this sequence repeats
in future data, this could be a new ER UMa system. The cyclic behavior of SOBs interspersed with normal outbursts in this small group of CVs is thought to be a combination of high mass transfer rate combined with a tidal
instability present in short period systems \citep{O96}.

SIMBAD identifies ZTF1752+07 with V982 Oph and classifies it as a Long Period Variable Candidate, but the ZTF
lightcurve (Figure 9) and the blue colors at
bright states are more consistent with a dwarf nova classification, as proposed by \citet{AS02}.

\section{Completeness}
At this time, it is difficult to obtain a good estimate of how complete the Marshal is in finding all CVs.
This first year was hampered by a late start of the Marshal filter, the loss of the month of October due
to equipment improvements, significant weather losses in the winter months, and the lack of good reference 
images for all the fields. In addition, the CVs in any field may not have had a dwarf nova outburst
or a change in low/high state during the months they were available in the sky. A rough estimate can be made by using the results from SDSS spectra and
the recent analysis of \citet{P19} who used the overlap between Gaia and SDSS colors to determine the
completeness of the CVs within 150 pc. The $\sim$300 available spectra from the SDSS Legacy program
provided a lower limit estimate of the density of CVs as 0.03 deg$^{-2}$ \citep{Sz03,Sz11}, and \citet{P19} estimated that the spectral completness of SDSS
is $\sim$57\%. ZTF covers about 62\% of the sky (25,774 deg$^{2}$ so this
should mean about 770 CVs should have been found, while only 308-329 were
(which would be about 40\% completeness). However, given the down time and weather, the
actual usable time in the survey is probably closer to 6-9 months (390-580
CVs) which would be more like 53-80\%. Since
the SDSS Legacy program was oriented toward high galactic latitudes to
sample quasars, larger numbers for the stellar density are expected with the inclusion of
low latitudes, so the lower
percentages are likely more realistic. A better estimate will be possible
as the survey proceeds and the lost months are reobserved and the reference
fields are completed.

\section{Conclusions}
Using the GROWTH Marshal to filter nightly alerts from ZTF $g$ and $r$ light curves throughout the first year of operation resulted in the identification of 90 known CVs, 218 strong candidates based on the shape, amplitude and colors of the
light curves, and an additional 21 potential candidates which require further data. Followup spectra obtained on a variety 
of 1.5-10 meter telescopes allowed spectroscopic confirmation of 27 of the 218 strong candidates from Balmer emission
lines, with an additional two with SDSS spectra. Unlike previous surveys, almost half of the new ZTF candidates are located within 10 degrees of the galactic plane,
demonstrating the capability of the ZTF camera and software to discover objects in crowded fields. While only 13\% of
the strong candidates have available and significant Gaia parallaxes, most
of their absolute magnitudes are consistent with
the faint end of the CV distribution (10-12), similar
to the CRTS. The outburst amplitudes increase with fainter absolute
magnitudes in this range, with many of those fainter than 12th being good
candidates for TOADs \citep{HSC95}. 

Four of the objects with spectra show high excitation \ion{He}{2} or only helium lines and deserve further time-resolved spectra to
determine their correct classification as either a system containing a magnetic white dwarf or an SW Sex system, or an AM CVn type.
Hour long time-resolved spectra of the known CV ZTF1631+69 shows strong \ion{He}{2} along with doubled Balmer emission lines,
implying an IP origin. A previously identified long period variable (V982 Oph) is more consistent with a dwarf nova
classification as proposed by \citet{AS02}.

While the available Marshal filter is not complete in finding all the CVs, it does demonstrate that even with several
ongoing surveys i.e. CRTS, ASASSN, MASTER, there are many systems being missed, especially those in the galactic
plane. Completeness in any ground survey is difficult to obtain due to weather and software ability in crowded fields. Additionally, followup
spectra of all the candidates is a time-consuming venture and will require increasingly large telescopes to obtain
spectra at quiescent magnitudes. Unfortunately, classification is best near quiescence when the emission lines produce the most information from the intensity and shape as to the correct type of CV. Some compromise can be reached by obtaining observations midway from outburst to decline.

\acknowledgments

PS and BD acknowledge funding from NSF grant AST-1514737.
A.Y.Q.H. is supported by a National Science Foundation Graduate Research Fellowship under Grant No.\,DGE-1144469. MC is supported by the David and Ellen Lee Prize Postdoctoral Fellowship at the California Institute of Technology. MLG acknowledges support from the DIRAC Institute in the Department of Astronomy at the University of Washington. The DIRAC Institute is supported through generous gifts from the Charles and Lisa Simonyi Fund for Arts and Sciences, and the Washington Research Foundation. We would like to thank occasional observers on the UW APO ZTF follow-up team, including: Brigitta Sip\H{o}cz, James Davenport, Daniela Huppenkothen, Dino Bekte\v{s}evi\'c, Gwendolyn Eadie, and Bryce T. Bolin. 
The authors thank the Observatoire de la C\^{o}te d'Azur for support.
This work was supported by the GROWTH project funded by the National Science Foundation under PIRE Grant No.\,1545949. Based on observations obtained with the Samuel Oschin Telescope 48-inch and the 60-inch Telescope at the Palomar Observatory as part of the Zwicky Transient Facility project.
ZTF is supported by the NSF under grant AST-1440341 and a collaboration including Caltech, IPAC, the Weizmann Institute for Science,the Oskar Klein Center at Stockholm University, the University of Maryland, the University of Washington, Deutsches Elektronen-Synchrotron and Humboldt University, Los Alamos National Laboratories, the TANGO Consortium of Taiwan, the University of Wisconsin at Milwaukee, and Lawrence Berkeley national Laboratories. Operations are conducted by COO, IPAC and
The SED Machine is based upon work supported by NSF under grant 1106171.
Some observations were made with the Apache Point 3.5m telescope, which is owned and operated by the Astrophysical Research Corporation. The Liverpool Telescope is operated on the island of La Palma by Liverpool John Moores University in the Spanish Observatorio del Roque de los Muchachos of the Instituto de Astrofisica de Canarias with financial support from the UK Science and Technology Facilities Council. The William Herschel Telescope
is operated on the island of La Palma by the Isaac Newton Group of Telescopes in the Spanish Observatorio del Roque de los Muchachos of the Instituto de Astrofisica de Canarias.

\vspace{5mm}
\facilities{APO:3.5m;Hale;ING:Herschel;Keck:I;Liverpool:2m;PO:1.2m;PO:1.5m}

\clearpage
\startlongtable 
\begin{longrotatetable}
\begin{deluxetable}{lccccccccccl}
\tabletypesize{\footnotesize}
\tablewidth{0pt}
\tablecaption{Known Confirmed CVs}
\tablehead{
\colhead{ZTF} & \colhead{RA} & \colhead{Dec} & \colhead{b$^{\circ}$} & \colhead{$\Delta$mag} & \colhead{p(mas)} & \colhead{Out} & \colhead{Days} &  \colhead{SDSS} & \colhead{CRTS} & \colhead{Spec\tablenotemark{a}} & \colhead{ID \& Other Surveys\tablenotemark{b}} }       
\startdata
  17aaaemzh &  00 15 38.27 &  +26 36 56.5 & -35.6 &  13.8-21.7 &   $1.64 \pm 0.19$ &  1   &  125 &   Y  &  Y3   & SD   &  AT2016eav,G \\
  18abdlywu &  00 38 54.83 &  +61 13 00.2 &  -1.6 &  14.3-20.5 &   $1.79 \pm 0.26$ &  2   &  224 &  --- &  ---  &  --- & KP Cas,G \\
  18abgopgb &  01 07 03.88 &  +42 43 12.0 & -20.1 &  15.5-20.0 &   $1.06 \pm 0.25$ &  SOB &  315 &   Y  & Y3    & AP   & IZ And,AT2018akr,G \\
  17aaawpsz &  02 13 15.49 &  +53 38 22.8 &  -7.3 &  16.8-20.0 &   ---             &   2  &  137 &  --- &  ---  &  --- & MOT,Gx,Atel5536,K,G \\
  18acgplgw &  02 50 00.20 &  +37 39 22.1 & -19.5 &  14.3-20.6 &   $1.99 \pm 0.11$ &   4  &  153 &  --- &  Y3   &  --- & PY Per,G  \\
  18aabeymw &  03 20 15.29 &  +44 10 59.1 & -11.0 &  15.3-20.9 &   $2.11 \pm 0.24$ & SOB  &  143 &   Y  &  ---  &  --- & USNO,K,G \\
  18abtmzoi &  04 06 59.82 &  +00 52 43.7 & -35.3 &  15.4-20.8 &   $1.78 \pm 0.20$ & 1    &  179 &   Y  & Y3    &  --- & CBET1463,G \\
  17aaaslud &  04 08 34.99 &  +51 14 48.1 &  -0.4 &  13.7-21.0 &   $1.71 \pm 0.05$ & 6    &  250 &   Y  &  ---  &  --- & FO Per,G \\
  17aaarlrs &  04 09 12.17 &  +48 22 06.2 &  -2.5 &  16.2-20.3 &   $1.11 \pm 0.11$ & 11  &  230 &  --- &  ---  &  --- & MY Per \\
  18abscxct &  04 23 32.91 &  +74 52 50.2 &  17.5 &  13.3-20.2 &   $3.52 \pm 0.11$ &  SOB &  223 &  --- &  ---  &  --- & HamburgSurvey,G \\
  17aaacdos &  04 53 16.88 &  +38 16 28.4 &  -3.6 &  15.0-19.0 &   $1.69 \pm 0.34$ &  7   &  233 &  --- &  ---  &  --- & HV Aur,G \\
  17aacucfy &  05 01 24.15 &  +20 38 17.7 & -12.9 &  15.8-20.2 &   $1.84 \pm 0.21$ &  5   &  163 &  --- &  Y3   &  AP  & ATel2266,G \\
  17aaavfwx &  05 06 13.12 &  -04 08 07.7 & -25.2 &  13.3-20.1 &   $2.66 \pm 0.12$ &  2   &  124 &   Y  &  Y3   &  --- & AQ Eri,G \\
  17aadaxwu &  05 47 48.38 &  +28 35 10.9 &   0.2 &  14.1-16.5 &   $1.83 \pm 0.07$ &  1   &    3 &  --- &  ---  &  --- & FS Aur,G  \\
  18abwsres &  05 58 45.48 &  +39 15 33.0 &   7.6 &  14.2-19.0 &   $1.81 \pm 0.25$ &SOB,1 &  237 &  --- &  ---  &  --- & USNO,AT2016ggz,K,G \\
  17aacpcfv &  06 12 04.47 &  +25 28 32.6 &   3.4 &  17.2-20.0 &   ---             &    2 &  327 &   Y  &  ---  & ---  & HQ Gem,AT2017gbf,G \\
  17aadnmap &  06 15 43.92 &  +28 35 08.4 &   5.6 &  16.6-20.0 &   $2.29 \pm 0.39$ &   HL &  68  &  --- &  ---  &  --- & KR Aur,G  \\
  17aacklbl &  06 16 43.23 &  +15 24 11.2 &  -0.5 &  13.2-20.0 &   $2.02 \pm 0.11$ &    8 &  541 &  --- &  ---  &  --- & CZ Ori,G \\
  18aaahnyx &  07 18 03.34 &  +64 47 44.7 &  27.2 &  15.3-18.2 &   $1.57 \pm 0.16$ &  1   &  217 &   Y  &  Y2   &  AP  & MOT,AT2018hzr,PTF,G \\
  18aagqdbp &  07 44 19.75 &  +32 54 48.2 &  24.8 &  17.8-20.3 &   ---             &  2   &  480 &   Y  &  Y4   &  --- & AM CVn type,G \\
  18aaicnwh &  07 50 59.97 &  +14 11 50.2 &  19.5 &  15.9-20.3 &   $1.32 \pm 0.28$ &  2   &  411 &   Y  &  Y3   &   SD & MOT,G \\
  17aacbiid &  07 51 07.52 &  +30 06 28.2 &  25.3 &  15.1-20.6 &   ---             &  2   &  300 &   Y  &  Y2   &  --- & K,G \\
  17aacplzj &  07 56 48.05 &  +30 58 05.0 &  26.7 &  16.8-20.4 &   ---             &  SOB &   87 &   Y  &  Y3   &   SD & G \\
  17aabyrpg &  08 08 46.20 &  +31 31 05.9 &  29.3 &  14.2-20.8 &   ---             &  2   &  556 &   Y  &  Y7   & SD   & PTF,G \\
  18aacluoi &  08 16 10.82 &  +45 30 10.1 &  33.4 &  15.9-20.3 &   ---             &  2   &  470 &   Y  &  Y3   & SD   & AAVSO,G \\
  18aaadlpa &  08 54 14.01 &  +39 05 36.7 &  39.8 &  16.2-19.3 &  $1.82 \pm 0.38$  &   HL &  472 &   Y  &  HL   & SD   & FR Lyn,G \\
  18aaacmsd &  09 43 25.89 &  +52 01 28.8 &  47.1 &  15.2-20.6 &  $1.21 \pm 0.26 $ &  5   &  346 &   Y  &  Y2   & SD   & G \\
  17aaapome &  10 05 15.37 &  +19 11 07.9 &  51.2 &  13.0-19.7 &  $2.57 \pm 0.24$  &  3   &  411 &   Y  &  Y2   & SD   & 2MASS,G \\
  18abcsatf &  10 18 12.99 &  +71 55 42.9 &  40.6 &  14.4-20.5 &   ---             &  3   &  326 &  --- &  ---  &  --- & CI UMa,G \\
  17aacldol &  10 19 47.24 &  +33 57 53.3 &  56.8 &  14.9-20.4 &  $1.37 \pm 0.25$  &  2   &  562 &   Y  &  Y3   &  SD  & AC LMi,G  \\
  18aaadhmx &  10 43 56.60 &  +58 07 31.4 &  51.8 &  15.8-20.8 &  ---              &  2   &  384 &   Y  &  ---  &  SD  & IY UMa \\
  17aaajlbs &  10 54 30.51 &  +30 06 09.1 &  64.2 &  14.3-18.7 &  $3.08 \pm 0.12$  &  4   &  562 &   Y  &  Y2   &  SD  & SX LMi,G \\
  17aaajlfw &  11 05 39.78 &  +25 06 28.0 &  66.2 &  14.0-19.3 &  $8.83 \pm 0.08$  & HL   &  532 &   Y  &    HL &  SD  & ST LMi,G  \\
  18aaadcme &  11 35 51.07 &  +53 22 46.2 &  60.3 &  15.1-20.6 &  ---              &  4   &  416 &   Y  &  ---  &  SD  & G \\
  18aaadclg &  11 57 44.85 &  +48 56 17.9 &  65.8 &  15.2-20.7 &  $0.60 \pm 0.03$  &  --- &  480 &   Y  &  Y    &  SD  & BE UMa(pre-CV)  \\
  18aajoejk &  12 27 40.85 &  +51 39 24.7 &  65.1 &  15.1-20.8 &  $2.75 \pm 0.20$  &  2   &  397 &   Y  &  Y4   &  SD  & USNO,G \\
  19aaqhcmn &  12 32 25.77 &  +14 20 41.7 &  76.5 &  13.4-20.6 &  ---              &  SOB &  103 &   Y  &  ---  &  --- & AL Com,G \\
  18aambkqd &  13 05 14.74 &  +58 28 56.2 &  58.6 &  16.9-20.9 &  ---              &  5   &  410 &   Y  &  Y3   &  SD  & G \\
  18aalrikz &  13 07 53.84 &  +53 51 30.3 &  63.1 &  15.8-20.9 &  $1.51 \pm 0.12$  &  HL  &  408 &   Y  &  Y4   &  SD  & EV UMa,G  \\
  18aautxxk &  14 11 18.31 &  +48 12 57.6 &  63.8 &  13.0-19.5 &  $2.36 \pm 0.30$  &  SOB &  410 &   Y  &  Y3   & K,AP & AM CVn type \\
  18abfyzmf &  14 25 48.07 &  +15 15 01.2 &  65.1 &  17.1-20.8 &  ---              &   2  &  324 &   Y  &  Y2   &  --- & K \\
  18aagsgqc &  14 57 44.75 &  +40 43 40.5 &  60.7 &  12.8-21.0 &  $1.47 \pm 0.20$  &   2  &  312 &  --- &  Y2   &  --- & TT Boo,G \\
  18abaulyr &  15 34 12.18 &  +59 48 31.8 &  47.2 &  14.8-20.0 &  $1.74 \pm 0.43$  &   1  &   31 &   Y  &  Y1   &  --- & DM Dra,G \\
  18aaisedb &  15 51 22.39 &  +71 45 11.6 &  39.1 &  15.0-19.8 &  $1.89 \pm 0.05$  &  18  &  409 &  --- &  ---  &  --- & SS UMi,G  \\
  18adbahiw &  15 56 44.23 &  -00 09 50.4 &  37.8 &  14.5-19.3 &  $3.24 \pm 0.21$  &   1  &  125 &   Y  &  Y3   &  SD  & V493 Ser,AT2018hbm,G \\
  18aaomiig &  16 00 03.71 &  +33 11 13.7 &  49.1 &  14.1-20.0 &  ---              & SOB,1&  389 &   Y  &  Y12  &  --- & VW CrB,G \\
  18abjzbhm &  16 04 19.02 &  +16 15 48.3 &  44.2 &  17.5-20.0 &  $1.10 \pm 0.27$  &   1  &  308 &   Y  &  Y5   &  SD  & MNRAS,G \\
  18aauxwft &  16 19 35.80 &  +52 46 31.6 &  44.0 &  16.3-21.0 &  $2.28 \pm 0.23$  &  HL  &  383 &   Y  &  Y2   &  SD  & 2MASS,G \\
  18aabpwzq &  16 22 07.16 &  +19 22 36.5 &  41.3 &  15.6-19.7 &  $1.01 \pm 0.17$  &   3  &  317 &   Y  &  ---  &  SD  & V589 Her,G \\
  18abaaewz &  16 31 00.23 &  +69 50 01.2 &  37.2 &  17.2-21.0 &  ---              &  HL  &  379 &  --- &  Y2   &  AP  & AT2018fmi,ROSAT,XMM,G \\
  18aaovjvr &  16 52 44.83 &  +33 39 25.5 &  38.3 &  18.1-20.1 &  ---              &   8  &  411 &   Y  &  Y1   &  SD  & G \\
  18aaiytds &  17 27 58.13 &  +38 00 22.5 &  32.0 &  14.6-20.9 &  $1.77 \pm 0.14$  &   4  &  389 &   Y  &  ---  &  MDM & MOT,G\\
  18aajtkma &  17 30 08.36 &  +62 47 54.3 &  33.2 &  15.6-20.0 &  ---              &   3  &  236 &   Y  &  ---  &  SD  & 2MASS,G \\
  18aalafea &  17 31 02.22 &  +34 26 33.1 &  30.7 &  16.3-20.7 &  ---              &  SOB &  408 &   Y  &  Y1   &  AP  & ASASSN-15cm,G \\
  18aakzxki &  17 42 09.17 &  +23 48 29.4 &  25.2 &  14.2-20.0 &  $1.26 \pm 0.16$  &   5  &  380 &   Y  &  Y3   &  --- & V660 Her,AT2018cyb,G\\
  18abetddh &  17 47 14.33 &  +15 00 47.5 &  20.8 &  17.1-20.0 &  ---              &   8  &  354 &   Y  &  Y2   &  --- & G \\
  18aajrvst &  17 48 16.33 &  +50 17 22.9 &  30.4 &  16.4-20.6 &  ---              &   11 &  386 &   Y  &  Y3   &  --- & AT2016bnf,MOT,G \\
  18aaslagi &  18 05 46.35 &  +31 40 17.7 &  22.9 &  14.1-20.7 &  $1.83 \pm 0.18$  &    2 &  365 &  --- &  Y4   &  --- & V1008 Her,G,AT2019akt \\
  18abbprmq &  18 08 44.69 &  +34 27 23.9 &  23.2 &  17.8-20.0 &  ---              & SOB,1&  101 &  --- &  ---  &  --- & V631 Her  \\
  18aamigoo &  18 16 13.17 &  +49 52 05.1 &  25.9 &  12.3-15.5 &  $11.40 \pm 0.02$ &  H,L &  379 &  --- &  Y2   &  --- & AM Her,G  \\
  18abdhozj &  18 32 11.38 &  +61 55 06.0 &  25.9 &  15.8-19.0 &  ---              &  2   &  339 &  --- &  ---  &  AP  & ASASSN-13ah,ATel5052,G \\
  18aakzfjo &  18 44 26.67 &  +37 59 51.9 &  17.6 &  13.5-20.0 &  $2.22 \pm 0.13$  &  12  &  382 &  --- &  ---  &  --- & AY Lyr,G,AT2019njr  \\
  18aakzafr &  18 44 39.18 &  +43 22 28.0 &  19.4 &  15.5-20.0 &  $0.94 \pm 0.08$  &  5   &  379 &  --- &  Y2   &  --- & V344 Lyr,G  \\
  18aapgtye &  18 53 00.76 &  +45 27 08.2 &  18.7 &  16.0-20.0 &  $0.86 \pm 0.19$  &  5   &  376 &  --- &  Y2   &  --- & KIC,G \\
  18abdkpgs &  19 14 43.53 &  +60 52 13.8 &  20.8 &  15.3-19.9 &  ---              &  1   &  336 &  --- &  ---  &  --- & CBET1535,AT2017eqn,K \\
  18aavetqn &  19 18 42.00 &  +44 49 12.4 &  14.3 &  15.5-21.0 &  $1.22 \pm 0.31$  &  2   &  228 &  --- &  ---  &  --- &KIC,Gx,ATel6187,AT2018fao,G \\
  18aaptcay &  19 19 05.19 &  +48 15 06.1 &  15.6 &  18.5-22.7 &  ---              &  9   &  378 &  --- &  ---  &  --- & AM CVn type,PTF1,G \\
  18abcccnr &  19 22 41.96 &  +52 43 59.0 &  16.8 &  14.0-21.0 &  ---              &  4   &  327 &  --- &  ---  &  --- & V1113 Cyg,G  \\
  18aauefbw &  19 27 48.53 &  +44 47 24.6 &  12.8 &  16.8-20.0 &  $1.16 \pm 0.17$  &  2   &  371 &  --- &  ---  &  --- & KIC,G \\
  18aaptcqq &  19 44 19.29 &  +49 12 57.3 &  12.3 &  18.2-20.0 &  ---              &  13  &  366 &  --- &  ---  &  --- & KIC, Gx \\
  18abmszln &  19 48 14.46 &  +34 52 01.1 &   4.7 &  16.4-20.8 &  ---              &  2   &  277 &  --- &  ---  &  --- & V1153 Cyg,G  \\
  18aasncio &  19 48 23.31 &  +36 26 23.0 &   5.4 &  14.0-20.0 &  $1.95 \pm 0.05$  &  9   &  361 &  --- &  ---  &  --- & V811 Cyg,G \\
  18aawbluo &  19 53 04.93 &  +21 14 48.8 &  -3.2 &  15.5-19.2 &  $0.62 \pm 0.05$  &  4   &  349 &  --- &  ---  &  --- & V405 Vul,G \\
  18abobptn &  19 57 18.83 &  -09 19 21.5 & -18.8 &  12.6-18.6 &  $3.19 \pm 0.06$  &  4   &  300 &   Y  &  Y2   &  --- & UU Aql,G \\
  18abjhdua &  19 58 37.08 &  +16 41 28.4 &  -6.6 &  14.9-20.6 &  $1.50 \pm 0.34$  &  1   &   76 &  --- &  ---  &  --- & AW Sge,G  \\
  18aawvwks &  19 59 52.40 &  +39 13 59.8 &   4.9 &  15.3-20.7 &  ---              &  2   &  347 &  --- &  ---  &  --- & V337 Cyg,G \\
  18aavzjuw &  20 00 05.22 &  +22 56 06.0 &  -3.7 &  17.0-20.0 &  $0.76 \pm 0.11$  &  5   &  350 &  --- &  ---  &  --- & SW Vul,G \\
  17aabopqx &  20 12 13.66 &  +42 45 50.9 &   4.8 &  16.3-19.7 &  $1.51 \pm 0.14$  &  4   &  342 &  --- &  ---  &  --- & V1316 Cyg,G  \\
  18aaypnnd &  21 08 33.97 &  +39 05 35.3 &  -5.8 &  15.7-18.4 &  $1.73 \pm 0.07$  &  6   &  332 &  --- &  ---  &  --- & Lanning386,G \\
  18aazsdnv &  21 26 24.12 &  +25 38 26.7 & -17.7 &  14.3-20.8 &  $1.88 \pm 0.33$  &  5   &  327 &   Y  &  ---  &  --- & MOT,AT2016gwu,ATel5111,K,G \\
  17aaaqgbm &  21 34 15.86 &  +49 11 26.3 &  -2.0 &  15.3-20.2 &  $1.83 \pm 0.12$  &  1   &  331 &  --- &  ---  &  --- & V1081 Cyg,G  \\
  17aaawglf &  21 44 03.78 &  +44 39 01.7 &  -6.5 &  15.7-20.0 &  $0.99 \pm 0.29$  &  10  &  328 &  --- &  ---  &  --- & V2209 Cyg,G \\
  18abcsuit &  21 47 38.41 &  +24 45 54.0 & -21.8 &  13.2-21.0 &  $2.01 \pm 0.22$  &  3   &  208 &   Y  &  Y3   &  --- & KIC,G \\
  18abqdtes &  21 54 33.97 &  +23 54 00.1 & -23.5 &  16.2-21.0 &  ---              &  2   &  106 &  --- &  ---  &  SM  & MOT,G \\
  17aaaedpn &  21 57 16.44 &  +52 12 00.8 &  -2.0 &  16.4-21.0 &  ---              &  7   &  333 &  --- &  ---  &  --- & V1404 Cyg,G  \\
  18abcqadc &  22 19 10.14 &  +31 35 22.8 & -21.0 &  17.2-20.4 &  ---              &  4   &  136 &   Y  &  Y2   &  --- & PTF1,G \\
  18abigrzf &  22 21 44.77 &  +18 40 07.9 & -31.6 &  13.3-19.0 &  $5.23 \pm 0.09$  &  3   &  317 &   Y  &  Y1   &   SD & V521 Peg \\
  18aaznfkp &  22 23 04.66 &  +52 40 58.2 &  -3.9 &  15.7-20.0 &  ---              &  3   &  328 &  --- &  ---  &  --- & MN Lac,G  \\
  18abtffxi &  22 24 43.46 &  +50 31 39.1 &  -5.8 &  16.1-20.9 &  ---              &  SOB &   41 &  --- &  ---  &  --- &  MR Lac \\
  18abccqjx &  22 43 40.73 &  +30 55 20.0 & -24.4 &  13.7-17.1 &  $1.36 \pm 0.04$  &  6   &  320 &   Y  &  Y3   &  --- & V537 Peg,G \\
\enddata
\tablenotetext{a}{AP=APO DIS, K=Keck LRIS, P=Pal200in DBS, SD=SDSS, SM=SEDM, SP=SPRAT}
\tablenotetext{b}{MOT=MASTEROT, G=Gaia, Gx=GALEX, KIC=Kepler, K=Kato SH papers}
\end{deluxetable}
\end{longrotatetable}

\clearpage
\startlongtable
\begin{longrotatetable}
\begin{deluxetable}{lccccccccccl}
\tabletypesize{\footnotesize}
\tablewidth{0pt}
\tablecolumns{12}
\tablecaption{CV Candidates}
\tablehead{
\colhead{ZTF} & \colhead{RA} & \colhead{Dec} & \colhead{b$^{\circ}$} & \colhead{$\Delta$mag} & \colhead{p(mas)} & \colhead{Out} & \colhead{Days} &  \colhead{SDSS} & \colhead{CRTS} & \colhead{Spec\tablenotemark{a}} &  \colhead{Other Surveys\tablenotemark{b}} }
\startdata
 18ablvaya &  00 14 36.26 &  +50 32 39.9 & -11.9 &  16.7-21.0 & ---             & SOB,2 & 147 & --- & --- & --- & MOT,Gx,ATel5749,G \\
 18abcrbtv &  00 14 00.93 &  +59 48 02.8 &  -2.7 &  18.0-20.0 & ---             &     3 & 237 & --- & --- & --- & G \\
 18abuocxn &  00 17 09.52 &  +48 53 49.6 & -13.6 &  15.1-21.0 & ---             &   SOB &  92 & --- & --- & --- & G \\
 18abgjgyt &  00 20 09.52 &  +51 19 13.0 & -11.2 &  17.6-20.6 & ---             &     4 & 235 & --- & --- & --- & G \\
 18abgrzbs &  00 33 03.96 &  +38 01 05.3 & -24.7 &  16.0-20.8 & $1.48 \pm 0.19$ &     3 & 226 &  Y  &  Y7 & --- & AT2016dtx,G \\
 17aabumcd &  00 34 17.70 &  +43 45 39.3 & -19.0 &  17.4-21.0 & ---             &   SOB & 154 & --- & --- & --- & G \\
 18aczower &  00 34 59.91 &  +27 36 19.0 & -35.1 &  18.3-20.4 & ---             &     3 & 264 &  Y  &  Y2 & SDe & G \\
 18accbuhi &  00 36 15.24 &  +38 51 31.8 & -23.9 &  18.5-23.0 & ---             &   SOB &  36 &  Y  &  Y2 &  SP & PTF \\
 18abmntno &  00 47 30.31 &  +58 32 01.8 &  -4.3 &  16.0-20.0 & ---             &     2 & 182 & --- & --- & --- & G \\
 17aaaeefu &  00 59 43.51 &  +64 54 42.2 &   2.1 &  16.3-18.7 & $1.32 \pm 0.07$ &     6 & 250 & --- & --- & --- & G \\
 18abgtrkj &  01 12 37.73 &  +61 27 35.4 &  -1.3 &  16.2-20.1 & $1.10 \pm 0.19$ &   SOB & 200 & --- & --- & --- & G \\
 18abvytad &  01 14 17.73 &  +59 35 54.7 &  -3.1 &  17.3-21.0 & ---             &   SOB &  41 & --- & --- & --- & --- \\
 18abumrmx &  01 15 16.55 &  +24 55 30.1 & -37.6 &  16.6-19.5 & ---             &     2 & 300 &  Y  &  Y3 & SDe & G \\
 18abflqfh &  01 17 12.39 &  +58 28 04.3 &  -4.2 &  16.5-19.6 & ---             &     8 & 234 & --- & --- & --- & G \\
 17aaaruaj &  01 28 36.51 &  +53 28 29.3 &  -9.0 &  18.0-21.0 & ---             &     7 & 233 & --- & --- & --- & MOT,Gx,ATel5395,G \\
 18abzbknn &  01 32 37.26 &  +60 36 09.3 &  -1.9 &  18.0-21.0 & ---             &     2 & 152 & --- & --- & --- & --- \\
 18abqasii &  01 34 38.24 &  +52 06 15.9 & -10.2 &  16.1-20.5 & ---             &     3 & 185 &  Y  & --- & --- & G \\
 18aabfagc &  02 02 31.50 &  +56 23 38.9 &  -5.1 &  16.2-20.0 & $0.69 \pm 0.14$ &     8 & 228 & --- & --- & --- & G \\
 18abtlqub &  02 04 58.23 &  +57 28 06.9 &  -4.0 &  17.6-20.0 & ---             &   SOB &  54 & --- & --- & --- & --- \\
 17aaatlsc &  02 14 12.97 &  +52 25 56.8 &  -8.4 &  18.3-20.7 & ---             &     6 & 341 & --- & --- & --- & --- \\
 18aabfbek &  02 45 07.91 &  +46 33 00.0 & -12.0 &  16.1-19.9 & ---             &     2 &  37 &  Y  & --- & --- & G \\
 18ablwmaq &  02 45 43.60 &  +63 16 48.7 &   3.2 &  17.3-20.1 & ---             &     3 & 224 & --- & --- & --- & G \\
 18abtmqvh &  03 32 24.87 &  +53 36 17.3 &  -2.1 &  16.5-20.0 & ---             &     2 & 207 & --- & --- & --- & G \\
 18acsjgoa &  03 42 46.18 &  +33 42 43.2 & -16.8 &  17.7-20.5 & ---             &     2 & 122 & --- &  Y3 & --- & G \\
 18adarypz &  04 01 17.42 &  +28 39 41.7 & -18.0 &  16.9-20.5 & ---             &   SOB &  45 & --- &  Y1 & --- & --- \\
 18acuwgij &  04 02 14.23 &  +22 08 35.5 & -22.5 &  17.5-21.0 & ---             &     1 &  30 & --- &  Y2 &  Pe & AT2018jrm \\
 19aactcry &  04 03 41.03 &  +47 56 26.4 &  -3.4 &  18.7-20.2 & ---             &   SOB &  47 & --- & --- & --- & --- \\
 18aabffpk &  04 22 15.21 &  +29 47 58.5 & -14.0 &  17.7-20.2 & ---             &     6 & 171 &  Y  &  Y3 & APe & G \\
 18aaadvfu &  04 28 37.10 &  +31 57 57.7 & -11.5 &  17.5-20.8 & ---             &     5 & 131 & --- &  Y2 & --- & G \\
 18achqmhh &  04 29 47.30 &  +18 47 48.4 & -20.0 &  17.9-20.1 & ---             &   SOB &  55 & --- & --- & APe & AT2018ild \\
 18acsjdxo &  04 39 16.69 &  +15 26 33.6 & -20.3 &  18.3-19.6 & ---             &     1 &  29 & --- &  Y2 & SPe & AT2018jkg \\
 17aactzul &  04 51 22.36 &  +16 10 19.5 & -17.5 &  18.8-20.8 & ---             &     1 &  11 & --- & --- & Ke  & G \\
 18aaadvfh &  04 55 27.72 &  +38 58 59.0 &  -2.8 &  18.0-20.5 & ---             &     3 & 157 & --- & --- & --- & --- \\
 18abvlkyj &  04 56 44.78 &  +50 45 32.8 &   4.8 &  18.0-21.0 & ---             &   SOB &  42 & --- & --- & --- & ---\\
 17aaabekg &  05 04 28.06 &  +44 46 37.5 &   2.1 &  17.6-20.2 & ---             &     4 & 251 & --- & --- & --- & G \\
 18abuafzd &  05 21 55.88 &  +38 52 15.1 &   1.3 &  16.0-21.0 & ---             &     1 &  85 & --- & --- & --- & G \\
 17aaatnzw &  05 24 04.33 &  +37 04 06.2 &   0.6 &  16.2-20.3 & ---             &     3 & 227 & --- & --- & --- & G \\
 18abvtosb &  05 28 55.68 &  +36 18 38.8 &   1.0 &  15.9-20.9 & ---             &     1 & 222 & --- & --- & --- & CBET,G \\
 17aabdvhz &  05 37 31.70 &  +50 15 21.4 &   9.8 &  17.5-19.0 & ---             &     4 & 226 & --- & --- & --- & G \\
 17aaczwtq &  05 53 33.12 &  +24 42 33.0 &  -0.7 &  17.3-20.7 & ---             &     2 &  97 & --- & --- & --- & G \\
 18abuqngq &  06 11 25.06 &  +22 21 17.1 &   1.7 &  16.8-20.0 & ---             &     2 & 229 & --- & --- & --- & PTF \\
 17aaatdec &  06 13 32.17 &  +06 57 07.6 &  -5.2 &  17.2-20.6 & ---             &     7 & 201 &  Y  & --- & --- & G \\
 18abyyhsf &  06 16 00.22 &  +21 22 29.9 &   2.2 &  18.1-20.0 & ---             &     1 &  31 & --- & --- & --- & --- \\
 17aadbpoq &  06 18 09.21 &  +27 38 24.4 &   5.6 &  15.6-20.0 & ---             & SOB,3 & 411 & --- & --- & --- & G \\
 17aacpcfj &  06 19 52.22 &  +24 20 58.6 &   4.4 &  15.1-18.9 & ---             &     1 &  71 &  Y  & --- & --- & G \\
 18aaaanuc &  06 21 23.26 &  +19 13 00.3 &   2.3 &  16.5-19.8 & ---             &     3 & 217 & --- & --- & --- & G \\
 18aabtvzf &  06 26 22.62 &  +19 46 32.6 &   3.6 &  16.6-20.6 & ---             &     3 & 469 & --- & --- & --- & G \\
 17aabyrom &  06 29 54.23 &  +25 39 11.8 &   7.0 &  16.2-20.9 & ---             &     3 & 336 &  Y  & --- & --- & MOT,ATel6027,G \\
 17aacmlmj &  06 31 27.16 &  +22 12 26.7 &   5.8 &  17.6-20.6 & ---             &   SOB &  39 & --- & --- & --- & --- \\
 18abypyap &  06 31 45.99 &  +09 54 45.2 &   0.2 &  16.8-20.1 & ---             &     2 & 109 &  Y  & --- & --- & --- \\
 18abztcib &  06 33 30.75 &  +59 18 47.8 &  20.9 &  15.5-20.8 & ---             &     2 & 218 & --- & --- &  SM & AT2018gxm \\
 19aakmorl &  06 37 33.01 &  -09 35 42.1 &  -7.4 &  11.3-19.3 & $2.72 \pm 0.44$ &     1 &  79 & --- & --- & --- & AT2019dhk,G \\
 18achtpcj &  06 41 21.30 &  +44 41 05.6 &  17.0 &  17.3-20.2 & ---             &   SOB &  36 &  Y  & --- & --- & AT2018ila \\
 17aaaoxxi &  06 49 01.96 &  +08 59 44.6 &   3.6 &  16.6-20.1 & ---             &     5 & 449 & --- & --- & --- & G \\
 17aabziqr &  06 52 13.34 &  +30 57 22.2 &  13.7 &  15.7-21.0 & ---             &     6 & 545 & --- & --- & --- & G \\
 19aaekpin &  06 54 04.39 &  -09 41 03.5 &  -3.8 &  16.3-20.5 & ---             &   SOB &  59 & --- & --- & --- & AT2019awv,ROSAT \\
 18acphaci &  06 55 50.84 &  -09 32 37.6 &  -3.4 &  17.5-20.6 & ---             &     6 & 150 & --- & --- & --- & G \\
 18acuxkzq &  07 04 37.59 &  +06 01 13.3 &   5.7 &  17.6-20.0 & ---             &   SOB &  36 & --- & --- & --- & AT2018jii \\
 18acpvmnu &  07 07 35.97 &  -09 12 41.7 &  -0.6 &  15.0-20.5 & ---             &     2 &  67 & --- & --- & --- & AT2018itx,G \\
 18acnnxrd &  07 11 28.27 &  -03 26 15.0 &   2.9 &  16.5-20.0 & ---             & SOB,1 & 117 & --- & --- &  SM & G \\
 18acuxvld &  07 15 12.98 &  -06 42 38.3 &   2.2 &  14.6-19.0 & ---             &   SOB &  31 & --- & --- & --- & G \\
 18acvvthl &  07 49 40.33 &  +07 15 56.0 &  16.2 &  17.3-19.9 & ---             &     1 &  16 &  Y  & --- & --- & MOT \\
 18acqxeba &  08 03 16.61 &  +66 11 10.0 &  31.9 &  15.0-17.9 & $0.87 \pm 0.25$ &     1 & 153 &  Y  & --- & --- & G \\
 18aabuaxp &  08 19 31.24 &  +21 33 38.0 &  28.5 &  17.4-20.2 & ---             &     3 & 508 &  Y  &  Y6 & SMe & G \\
 18aabjjuj &  09 11 47.01 &  +31 51 01.8 &  42.5 &  14.8-20.5 & ---             & SOB,1 & 134 &  Y  &  Y4 & --- & G \\
 18acnocdo &  09 14 42.69 &  +67 10 36.9 &  38.5 &  15.0-20.1 & $1.12 \pm 0.33$ &   SOB & 181 &  Y  &     & APe & G \\
 18aczejci &  09 26 20.42 &  +03 45 42.4 &  35.8 &  17.6-20.5 & ---             &     6 & 144 &  Y  &  Y2 & --- & G \\
 18aabywzu &  10 37 38.65 &  +12 42 50.1 &  55.6 &  16.6-20.3 & ---             &     1 &  84 & --- &  Y2 & SMe & AT2016ags \\
 17aaclmhw &  10 53 33.76 &  +28 50 35.7 &  64.0 &  17.9-19.9 & ---             &     3 & 111 &  Y  & --- &K,We & --- \\
 19aacyjjz &  11 10 18.95 &  -05 22 18.4 &  49.3 &  16.4-20.0 & ---             &     1 & 171 & --- &  Y2 & --- & ATel1272,G \\
 18aaqepuc &  11 37 08.67 &  +51 34 50.9 &  61.8 &  17.4-20.8 & ---             &     3 & 375 &  Y  &  Y4 & --- & G \\
 18abbghiz &  12 56 09.84 &  +62 37 04.4 &  54.5 &  16.5-21.5 & ---             &   SOB &  43 &  Y  & --- & --- & MOT,ATel8846 \\
 18aabqewr &  12 58 32.22 &  +26 01 06.0 &  88.1 &  16.0-21.0 & $1.14 \pm 0.05$ &     1 & 511 &  Y  & --- & --- & ASASSN-18cr \\
 18aabxycb &  13 15 14.42 &  +42 47 44.6 &  73.6 &  16.9-20.5 & ---             &     9 & 511 &  Y  &  Y8 &APe,We& PTF,G \\
 18aajlfdq &  13 56 42.38 &  +61 30 24.4 &  53.9 &  16.4-21.4 & ---             &     2 & 387 &  Y  & --- & --- & ASASSN-13ap,ATel5118,G \\
 18aawqkva &  15 44 28.10 &  +33 57 26.4 &  52.4 &  17.1-24.0 & ---             &     3 & 410 &  Y  &  Y2 & --- & AT2018fhi,PTF \\
 18aavojbe &  15 46 52.71 &  +37 54 14.9 &  51.9 &  16.0-21.1 & ---             &     2 & 350 &  Y  &  Y4 & --- & PTF,G \\
 18abklaip &  15 50 30.38 &  -00 14 17.4 &  39.0 &  18.1-19.5 & ---             &     2 & 275 &  Y  &  Y2 & --- & G \\
 18aakuohk &  16 17 00.94 &  +62 00 24.6 &  41.6 &  14.7-20.5 & $2.03 \pm 0.08$ &    10 & 384 &  Y  &  Y2 & --- & PTF,G \\
 18abagclj &  16 26 05.66 &  +22 50 43.5 &  41.5 &  18.3-20.6 & ---             & SOB,1 & 356 &  Y  &  Y3 & --- & CRTS \\
 18aakvnlw &  16 47 48.00 &  +43 38 45.0 &  40.2 &  19.4-20.4 & ---             &     7 & 174 & --- &  Y1 &  We & G \\
 18abkikam &  17 04 44.51 &  +26 20 28.5 &  34.1 &  19.5-20.7 & ---             &     1 &  31 &  Y  & --- &SM,Ke& --- \\
 18aaisdps &  17 05 15.33 &  +72 44 03.2 &  33.6 &  16.5-20.8 & ---             &     6 & 309 &  Y  & --- & --- & MOT,G\\
 18abpaake &  17 06 06.10 &  +25 51 53.2 &  33.7 &  16.4-20.3 & ---             &     1 &  32 &  Y  & --- & --- & --- \\
 19aagdvdi &  17 11 38.40 &  +05 39 50.9 &  24.7 &  15.2-20.1 & ---             &     2 & 106 & --- & --- & --- & PTF,AT2019ath,G \\
 18abeechv &  17 16 02.90 &  +29 27 36.5 &  32.5 &  18.8-20.0 & ---             &   SOB &  45 &  Y  & --- &  Pe & --- \\
 18abfwukx &  17 26 24.11 &  +36 25 06.3 &  32.0 &  19.2-21.0 & ---             &     1 &  37 & --- & --- &  Ke & AT2018eky\\
 19aanvbqa &  17 27 50.17 &  +23 52 47.5 &  28.4 &  14.8-20.0 & ---             &   SOB &  64 &  Y  & --- & --- & AT2019cni \\
 18aapqotx &  17 32 30.30 &  +50 09 32.6 &  32.9 &  17.5-20.0 & ---             &  2SOB & 385 & --- &  Y2 & --- & PTF,G \\
 18aakzqjh &  17 46 48.85 &  +19 47 44.5 &  22.8 &  18.0-21.0 & ---             &    10 & 360 &  Y  &  Y2 &  --- & G \\
 18aaploaw &  17 47 25.61 &  +63 02 47.9 &  31.2 &  18.0-20.4 & ---             &     3 & 409 &  Y  & --- &  --- & AT2018eeb,ATLAS18spw \\
 18aajrtmo &  17 48 27.86 &  +50 50 39.7 &  30.4 &  14.9-20.5 & $1.35 \pm 0.25$ &     5 & 408 &  Y  &  Y1 &  --- & ASASSN-13ak,PTF,G\\
 18abckxfb &  17 49 21.78 &  +19 44 22.9 &  22.2 &  18.4-21.0 & ---             &  SOB &   47 & --- & --- &   SM & AT2016cya \\
 18abffzyg &  17 52 38.47 &  +07 33 04.5 &  16.5 &  15.8-20.5 & ---             &    4 &  261 & --- & --- &  --- & V982 Oph,G\\
 18abfmuvj &  17 53 30.49 &  +20 38 07.1 &  21.7 &  17.8-21.0 & ---             &    2 &  329 & --- & --- &   SM & AT2018dyr \\
 18aasfpxf &  17 56 33.39 &  +57 29 26.6 &  29.9 &  17.5-21.0 & ---             &    2 &  384 &  Y  & --- &  --- & G \\
 18aabtvdf &  17 56 39.55 &  +44 40 12.4 &  28.1 &  15.3-21.0 & $0.72 \pm 0.06$ &   22 &  426 & --- & --- &  --- & G \\
 18aaumvgk &  18 00 18.30 &  +15 15 28.4 &  18.1 &  16.5-20.0 & ---             &    1 &  279 & --- & --- &SM,APe& MOT,AT2018bsp,ATEL9204\\
 18aapauoa &  18 00 43.08 &  +21 01 34.2 &  20.2 &  17.0-20.6 & ---             &    6 &  382 & --- & --- &  --- & G \\
 18aagrotg &  18 02 31.35 &  +30 58 29.1 &  23.2 &  16.6-21.0 & ---             &    5 &  388 & --- & --- &  --- & G \\
 18abcmsux &  18 03 24.75 &  +17 52 31.6 &  18.4 &  18.2-21.0 & ---             &2SOB,1&  315 & --- & --- &  --- & AT2018fpf \\
 18aayzgkc &  18 10 20.69 &  +32 39 13.8 &  22.3 &  16.4-22.0 & ---             &  SOB &   73 & --- & --- &SM,APe& --- \\
 18abjvmhv &  18 15 27.34 &  +41 56 05.3 &  24.1 &  13.8-21.0 & ---             &  SOB &  299 &  Y  & --- &  --- & G \\
 18aaptdcp &  18 21 22.52 &  +61 48 55.2 &  27.2 &  14.3-21.1 & $1.15 \pm 0.29$ &    3 &  373 & --- & --- &  --- & ASASSN-13at,G \\
 18aaqznkg &  18 29 13.38 &  +31 06 23.6 &  18.0 &  19.5-21.8 & ---             &    1 &    9 & --- & --- &SM,Ke & MOT, AT2018bhx \\
 19aaedxnl &  18 30 07.21 &  +76 43 13.8 &  27.6 &  16.2-20.5 & ---             &    2 &  114 & --- & --- &  --- & MOT,ATel4843,G\\
 18aakytes &  18 31 20.68 &  +30 19 34.9 &  17.3 &  16.1-21.0 & ---             &    5 &  390 & --- & --- &  --- & MOT,ATel4761 \\
 18acaqbgu &  18 34 35.96 &  +31 32 00.9 &  17.1 &  17.6-21.0 & ---             &    1 &   30 & --- & --- &  --- & AT2016hrt,XMM,ROSAT \\
 18abfgyjt &  18 34 59.52 &  +54 33 15.0 &  24.1 &  17.9-21.0 & ---             &    2 &   31 & --- & --- &   Pe & --- \\
 18abdklug &  18 37 05.92 &  +37 17 58.8 &  18.7 &  18.5-20.0 & ---             &    2 &  336 & --- & --- &  --- & AT2018haz \\
 18aatzhmn &  18 41 51.05 &  +37 52 28.6 &  18.0 &  14.3-20.0 & ---             &  SOB &   78 & --- & --- &  --- & AT2018blk,G \\
 18aaracvu &  18 41 59.71 &  +37 34 11.6 &  17.9 &  17.8-20.0 & ---             &SOB,6 &  106 & --- & --- &K,APe & AT2018bit \\
 18acdwdgx &  18 42 51.29 &  +33 56 49.5 &  16.4 &  16.8-19.8 & ---             &    1 &   30 & --- & --- &  --- & G \\
 18aammzjo &  18 44 26.89 &  +36 51 40.2 &  17.2 &  17.8-20.6 & ---             &    12 & 410 & --- & --- &  --- & MOT,ATel6003,G \\
 18aavyoqk &  18 45 03.01 &  +13 55 17.6 &   7.7 &  14.9-18.9 & ---             &     6 & 350 & --- & --- &  --- & ROSAT,G \\
 18abfxhyn &  18 46 12.88 &  +12 52 29.0 &   7.0 &  17.2-20.8 & ---             &   SOB & 328 & --- & --- &  --- & MOT \\
 18ablrvfh &  18 47 15.37 &  -06 13 21.0 &  -1.9 &  16.8-19.6 & ---             &     4 & 273 & --- & --- &  --- & ROSAT,G \\
 18abcysbr &  18 48 43.45 &  +29 54 51.1 &  13.7 &  16.4-19.0 & ---             & SOB,2 & 279 & --- & --- &  --- & MOT \\
 18abixdpa &  18 50 22.30 &  +74 56 02.5 &  26.2 &  17.1-20.4 & ---             &     3 & 304 & --- & --- &  --- & --- \\
 18aawadah &  18 52 16.40 &  +15 06 23.5 &   2.1 &  16.4-20.5 & ---             &     3 & 377 & --- & --- &  --- & G \\
 18abbtilo &  18 57 44.62 &  +32 13 46.5 &  12.9 &  18.3-21.0 & ---             &   SOB &  38 & --- & --- &  APe & PTF \\
 18aaovfjr &  18 58 11.15 &  +46 27 56.5 &  18.2 &  15.4-20.4 & $1.42 \pm 0.28$ &     6 & 376 & --- &  Y1 &  --- & AT2017dac,G \\
 18abqazwf &  19 00 32.27 &  +30 30 40.0 &  11.6 &  17.9-20.0 & ---             &     2 & 270 & --- & --- &   Pe & --- \\
 18aaraifg &  19 00 58.43 &  +30 35 47.5 &  11.6 &  17.6-21.0 & ---             & SOB,1 &  33 & --- & --- &SM,Pe & AT2018bhy \\
 18accntcd &  19 01 04.62 &  +45 05 16.0 &  17.2 &  17.1-21.0 & ---             & SOB,1 & 201 & --- & --- &  --- & AT2018ijr \\
 18aapuklg &  19 03 42.18 &  +53 47 45.8 &  19.8 &  15.3-21.0 & $0.99 \pm 0.15$ & SOB,11& 378 & --- &  Y6 &  --- & G \\
 18abjkhgu &  19 05 16.00 &  +47 23 34.5 &  17.4 &  17.1-20.7 & ---             &   SOB & 288 & --- & --- &  --- & MOT,Atel9104\\
 18abciqza &  19 08 12.64 &  +04 57 28.0 &  -1.5 &  17.3-20.1 & $0.88 \pm 0.14$ &     3 & 319 & --- & --- &  --- & G \\
 18abcyzxp &  19 08 59.31 &  +25 39 45.7 &   7.8 &  17.6-19.1 & ---             &     1 &  10 & --- & --- &  --- & G \\
 18aavesgh &  19 14 04.90 &  +47 25 10.0 &  16.0 &  16.1-20.7 & ---             &     2 & 374 & --- & --- &  --- & G \\
 18aayupbw &  19 20 30.90 &  +27 43 33.1 &   6.5 &  17.4-21.0 & ---             &     1 & 280 & --- & --- &  --- & G \\
 18aapsxwc &  19 20 58.71 &  +45 06 37.6 &  14.0 &  18.2-20.0 & ---             &     6 & 373 & --- & --- &  --- & --- \\
 18aasnlwa &  19 22 34.05 &  +27 32 01.1 &   6.0 &  16.5-20.9 & ---             &     7 & 371 & --- & --- &  --- & G \\
 18abkhgfg &  19 26 33.86 &  +16 02 20.3 &  -0.3 &  16.9-20.3 & ---             &   SOB &  42 &  Y  & --- &  --- & AT2016hkw  \\
 18aayncoh &  19 27 57.59 &  +46 43 32.3 &  13.6 &  17.6-20.8 & ---             &     4 & 367 & --- & --- &  --- & --- \\
 18abdeelv &  19 30 59.19 &  +51 57 35.2 &  15.4 &  14.8-22.0 & ---             &   SOB & 327 & --- & --- &  --- & ASASSN-14fu,ATel6761,G \\
 18aauegwi &  19 32 03.61 &  +45 07 59.2 &  12.3 &  16.0-20.0 & $1.14 \pm 0.14$ &     5 & 373 & --- & --- &  --- & G \\
 18abptyvl &  19 35 57.46 &  +11 05 28.2 &  -4.6 &  15.6-20.8 & $1.38 \pm 0.43$ &     2 & 282 & --- & --- &  --- & G \\
 18abdkwtr &  19 35 57.54 &  +33 00 49.4 &   6.0 &  18.4-20.0 & ---             &     3 & 322 & --- & --- &  --- & ---  \\
 18abbctvx &  19 36 02.88 &  +46 08 31.2 &  12.1 &  17.1-20.0 & ---             &     1 &  31 & --- & --- &   SM & --- \\
 18abdiirq &  19 41 04.86 &  +31 57 47.7 &   4.5 &  15.4-21.0 & $0.71 \pm 0.06$ &     8 & 309 & --- & --- &  --- & G \\
 18aaynycd &  19 42 30.55 &  +36 01 19.3 &   6.3 &  18.3-20.1 & ---             &   SOB &  39 &  Y  & --- &   SM & --- \\
 18aazvbyj &  19 43 32.98 &  +16 31 06.8 &  -3.6 &  17.5-19.5 & ---             &     5 & 319 & --- & --- &  --- & G \\
 18abdkpol &  19 43 41.51 &  +43 59 11.9 &   9.7 &  18.2-20.5 & ---             &     2 & 114 & --- & --- &  --- & --- \\
 18aaptabw &  19 47 30.06 &  +48 09 17.6 &  11.3 &  16.1-21.2 & ---             &     3 & 358 & --- & --- &  --- & PTF \\
 18abfhssx &  19 48 26.68 &  +59 33 10.3 &  16.4 &  19.0-21.0 & ---             &     2 & 145 & --- & --- &   SM & \\
 18aauebur &  19 48 49.10 &  +46 30 36.1 &  10.3 &  18.5-21.0 & ---             & SOB,4 & 193 & --- & --- &   SM & AT2018epr \\
 18ablxhlc &  19 48 58.30 &  +58 26 20.9 &  15.8 &  18.2-21.0 & ---             &   SOB & 300 & --- & --- &  --- & G \\ 
 18abmfxml &  19 50 42.77 &  +08 25 45.3 &  -9.1 &  16.8-19.7 & ---             &     6 & 265 & --- & --- &  --- & G \\
 18abchfow &  19 50 38.85 &  +21 16 51.8 &  -2.7 &  18.6-21.0 & ---             &     3 & 349 & --- & --- &  --- & --- \\
 18aavzkpp &  19 50 54.38 &  +35 51 35.9 &   4.7 &  17.7-21.0 & ---             &     3 & 211 &  Y  & --- &  --- & AT2019bcl,G \\
 18aazfbvg &  19 54 46.08 &  +18 37 57.3 &  -4.8 &  16.2-17.6 & ---             &     8 & 330 & --- & --- &  APe & PTF,G \\
 18aaypfws &  19 55 37.25 &  +23 35 57.9 &  -2.5 &  19.0-21.0 & ---             &     4 & 349 & --- & --- &  --- & G \\
 18aayeczt &  19 57 12.21 &  +18 49 41.0 &  -5.2 &  16.5-19.0 & ---             &     2 & 174 & --- & --- &  --- & G \\
 18abvokyy &  19 57 15.07 &  +36 27 03.3 &   3.9 &  18.0-21.0 & ---             &     2 &  89 & --- & --- &  --- & G \\
 18aawlljc &  20 03 34.02 &  +29 34 36.7 &  -0.8 &  17.4-20.7 & ---             &     2 & 369 & --- & --- &   SM & AT2018ccf,G \\
 18abmhjpr &  20 04 07.64 &  +12 26 18.4 & -10.0 &  16.4-19.5 & $0.99 \pm 0.22$ &     5 & 272 &  Y  & --- &  --- & G \\
 18abnudvw &  20 04 20.95 &  +11 32 47.2 & -10.5 &  18.5-20.0 & ---             &     4 & 268 &  Y  & --- &  --- & AT2016diu,G \\
 18acyjvhm &  20 06 16.59 &  +07 38 15.4 & -12.8 &  15.9-19.8 & ---             &   SOB &  18 & --- & --- &  --- & --- \\
 18absihoy &  20 07 42.61 &  +11 58 10.4 & -10.9 &  16.5-20.0 & ---             &     2 & 140 & --- & --- &  --- & --- \\
 18abudsie &  20 07 44.61 &  -00 27 50.1 & -16.7 &  18.1-19.8 & ---             &     3 & 266 &  Y  & --- &  --- & G \\
 18abaqnny &  20 09 22.90 &  +59 13 48.1 &  13.9 &  19.2-24.0 & ---             &     2 &  70 &  Y  & --- &    K & --- \\
 18aayukga &  20 10 02.01 &  +37 06 15.4 &   2.1 &  17.5-20.8 & ---             &     6 & 329 & --- & --- &  --- & G \\
 18aaxyrau &  20 10 53.41 &  +44 10 46.3 &   5.8 &  17.4-19.7 & ---             & SOB,5 & 338 & --- & --- &  --- & MOT,ATel4611 \\
 18aawluad &  20 12 25.84 &  +23 36 20.9 &  -5.7 &  18.5-21.0 & ---             &     3 & 348 & --- & --- &  --- & G \\
 18abnuekr &  20 13 55.88 &  -05 26 06.1 & -20.8 &  17.2-19.5 & ---             &     3 & 259 & --- & --- &  --- & G \\
 18aayeosh &  20 26 23.46 &  +32 01 08.4 &  -3.6 &  15.7-21.0 & ---             &     5 & 346 & --- & --- &  --- & G \\
 18aazffjy &  20 28 36.83 &  +56 13 46.3 &  10.1 &  18.2-21.0 & ---             & SOB,4 & 342 & --- & --- &   SM & --- \\
 18abajlni &  20 32 32.12 &  +58 09 28.8 &  10.8 &  16.7-21.0 & ---             & SOB,2 & 286 &  Y  & --- &   SM & --- \\
 18abssrbs &  20 35 01.48 &  -00 19 46.4 & -23.0 &  15.6-20.0 & $1.17 \pm 0.25$ &     1 & 263 &  Y  &  Y3 &  --- & ASASSN-14hb,AT2019gya,G \\
 18aaxprrt &  20 36 39.95 &  +47 02 15.8 &   3.7 &  17.8-21.0 & ---             &     3 & 316 & --- & --- &  --- & G \\
 18acbwqtc &  20 40 44.21 &  +50 59 40.2 &   5.6 &  16.0-21.0 & ---             &   SOB &  39 & --- & --- &  --- & --- \\
 18abikzev &  20 41 30.41 &  +48 13 33.0 &   3.8 &  18.6-20.5 & ---             &     3 & 319 & --- & --- &  --- & G  \\
 18abdhxvl &  20 43 59.58 &  +42 03 25.4 &  -0.4 &  16.8-19.5 & $1.49 \pm 0.45$ &     6 & 335 &  Y  & --- &  --- & AT2016dvk,G \\
 18abxkbzz &  20 44 28.70 &  +14 50 11.3 & -16.9 &  18.4-21.0 & ---             &   SOB &  39 &  Y  & --- &   SM & AT2018grl \\
 18abzvpif &  20 49 51.90 &  +46 19 50.2 &   1.5 &  18.5-20.5 & ---             &     1 & 249 & --- & --- &  --- & --- \\
 18abjwsby &  20 59 15.71 &  +43 01 07.2 &  -1.9 &  17.4-20.0 & $0.45 \pm 0.12$ &     3 & 308 & --- & --- &  --- & G \\
 18aazeyze &  21 04 35.46 &  +15 05 01.4 & -20.7 &  18.0-20.3 & ---             &     4 & 329 & --- & --- &  --- & G \\
 18abmarba &  21 21 08.93 &  +30 34 14.3 & -13.5 &  16.0-20.8 & ---             &     2 & 287 & --- & --- &  --- & G \\
 17aaaejau &  21 21 28.58 &  +29 15 41.2 & -14.5 &  17.5-20.7 & ---             &     5 & 293 & --- & --- &  --- & AT2018kgn,G \\
 18absanfq &  21 23 05.54 &  +15 08 48.6 & -24.1 &  17.8-21.0 & ---             &     1 &  42 & --- & --- &   SM & AT2018fpd \\
 18aayoxjs &  21 25 43.95 &  +63 23 18.0 &   9.2 &  17.6-19.9 & ---             & SOB,5 & 334 & --- & --- &   SM & AT2018efi,G \\
 18abihypg &  21 28 22.20 &  +63 25 57.2 &   9.0 &  15.0-20.0 & ---             &     6 & 320 & --- & --- &   We & G \\
 18abvigco &  21 29 59.49 &  +78 52 21.5 &  19.7 &  14.8-20.0 & $1.89 \pm 0.32$ &   SOB &  45 & --- & --- &  --- & G \\
 18abwntpz &  21 34 37.09 &  +57 02 07.3 &   3.8 &  14.9-20.6 & ---             &   SOB & 220 & --- & --- &  --- & G \\
 18abaphmn &  21 38 25.68 &  +27 30 36.0 & -18.4 &  18.9-21.0 & ---             &   SOB & 102 &  Y  & --- &  --- & AT2016fvd \\
 18abjtach &  21 43 32.98 &  +46 12 22.3 &  -5.2 &  16.7-20.5 & ---             &   SOB &  33 & --- & --- &  --- & G \\
 17aadsnyp &  21 43 33.57 &  +51 20 03.7 &  -1.3 &  16.9-20.9 & $0.52 \pm 0.13$ &     7 & 331 & --- & --- &  --- & G \\
 17aaarine &  21 46 48.25 &  +48 27 47.5 &  -3.9 &  18.1-21.0 & ---             & 2 SOB & 122 &  Y  & --- &  --- & AT2016jbf \\
 18abnxgmb &  21 48 07.80 &  +29 38 48.0 & -18.3 &  16.4-21.0 & ---             &   SOB & 163 & --- & --- &  --- & G \\
 18abochac &  21 51 25.83 &  +46 28 03.5 &  -5.9 &  18.2-22.0 & ---             &     2 & 174 &  Y  & --- &  --- & G \\
 18abvxdou &  21 52 32.31 &  +49 04 19.4 &  -4.0 &  16.7-21.0 & ---             & SOB,1 & 263 & --- & --- &  --- & AT2018glg \\
 17aabutiy &  21 58 15.94 &  +51 50 20.3 &  -2.4 &  17.4-19.8 & ---             &     3 & 336 & --- & --- &  --- & G \\
 18abvbqyo &  22 06 41.05 &  +30 14 35.8 & -20.4 &  18.4-21.0 & ---             &     2 & 234 &  Y  &  Y1 &   SM & AT2016jai \\
 18acarunx &  22 14 03.77 &  +37 34 52.8 & -15.5 &  17.5-20.0 & ---             &     2 & 254 & --- & --- &  --- & AT2018hgp \\
 18abmnbne &  22 17 31.66 &  +46 59 32.9 &  -8.2 &  17.3-20.4 & ---             &   SOB &  67 & --- & --- &  --- & --- \\
 18ablvntg &  22 27 07.17 &  +55 38 00.0 &  -1.7 &  15.2-20.6 & $1.60 \pm 0.39$ &     1 &  41 & --- & --- &  SMe & G \\
 18abqboud &  22 31 23.00 &  +33 30 57.1 & -20.6 &  15.5-21.0 & ---             &   SOB & 108 & --- & --- &   SM & AT2018frz \\
 18abblwnw &  22 45 00.98 &  +56 31 30.6 &  -2.2 &  16.8-20.3 & ---             &     4 & 328 & --- & --- &  --- & G \\
 18actxlqb &  22 45 05.38 &  +01 15 47.2 & -48.4 &  18.4-21.6 & ---             &     1 &  28 &  Y  &  Y4 &  --- & AT2018jlx \\
 18abmnmuw &  22 47 51.45 &  +36 43 19.3 & -19.8 &  15.7-20.4 & ---             &     1 &  41 &  Y3 & --- &  --- & MOT,ATel7438,G \\
 18acehgym &  22 48 25.98 &  +38 55 09.6 & -18.0 &  19.2-23.0 & ---             &     2 &  32 & --- & --- &SM,Ke & AT2018hgz \\
 18abccwds &  22 52 31.58 &  +59 11 35.1 &  -0.2 &  18.9-21.0 & ---             &     4 & 364 & --- & --- &  --- & G \\
 18abmnmvz &  22 53 50.55 &  +33 30 32.4 & -23.3 &  16.2-20.2 & ---             &     3 & 159 &  Y  &  Y2 &  --- & ASASSN-13by,G \\
 18abvburn &  22 55 21.40 &  +53 38 43.8 &  -5.4 &  18.7-20.9 & ---             &     1 &  40 & --- & --- &  --- & ---  \\
 18ablvjse &  22 56 32.43 &  +35 42 38.4 & -21.6 &  17.0-21.0 & ---             &     3 & 297 & --- &  Y2 &  APe & MOT,G \\
 18abcxmen &  22 56 57.90 &  +55 53 41.6 &  -3.5 &  17.6-21.0 & ---             &     7 & 197 & --- & --- & ---  & --- \\
 18aaypqid &  23 05 38.42 &  +65 21 58.6 &   4.7 &  14.7-20.5 & $1.62 \pm 0.23$ &     1 & 332 & --- & --- & ---  & ROSAT,G \\
 18aazndjw &  23 10 11.15 &  +51 15 11.1 &  -8.5 &  17.6-20.5 & ---             &   SOB & 329 & --- & --- &  SM  & ATel11797,AT2018ctl,ROSAT,G \\
 18aazwddf &  23 24 04.01 &  +51 43 18.0 &  -8.8 &  17.3-21.4 & ---             &     4 & 328 &  Y  & --- & ---  & --- \\
 17aabunmx &  23 34 35.56 &  +54 33 25.5 &  -6.7 &  15.7-20.5 & $1.19 \pm 0.19$ & SOB,3 & 225 & --- & --- & ---  & AT2018asi,G \\
 18abvxbfh &  23 35 30.00 &  +60 54 05.7 &  -0.6 &  17.4-20.3 & ---             &     1 &  52 & --- & --- & ---  & --- \\
 17aabuvei &  23 36 46.10 &  +57 57 24.1 &  -3.5 &  17.7-20.0 & ---             &     7 & 231 & --- & --- & ---  & G \\
 18abcpbbj &  23 37 27.57 &  +51 13 58.7 & -10.0 &  16.7-21.0 & $1.71 \pm 0.33$ &     3 & 206 & --- & --- & ---  & G \\
 18abjmxql &  23 38 43.54 &  +57 17 19.9 &  -4.2 &  15.2-20.4 & ---             &     2 & 143 &  Y  & --- & ---  & G \\
 18abiwzlg &  23 52 01.18 &  +44 50 58.2 & -16.8 &  15.2-20.4 & $1.60 \pm 0.26$ & SOB,1 & 178 & --- &  Y1 & ---  & AT2016,G \\
 17aaaedem &  23 54 58.63 &  +54 27 29.3 &  -7.5 &  17.0-21.0 & ---             & SOB,5 & 205 & --- & --- & ---  & --- \\
 18abgtjea &  23 59 33.64 &  +56 05 01.5 &  -6.1 &  17.6-21.0 & ---             &   SOB &  65 & --- & --- & ---  & AT2018eoi,ATLAS \\
\enddata
\tablenotetext{a}{AP=APO DIS, K=Keck LRIS, P=Pal200in DBS, SD=SDSS, SM=SEDM, SP=SPRAT, W=WHT}
\tablenotetext{b}{MOT=MASTEROT, G=Gaia, Gx=GALEX}
\end{deluxetable}
\end{longrotatetable}

\clearpage
\startlongtable
\begin{longrotatetable}
\begin{deluxetable}{lccccccccccl}
\tabletypesize{\footnotesize}
\tablewidth{0pt}
\tablecolumns{12}
\tablecaption{Possible CV Candidates}
\tablehead{
\colhead{ZTF} & \colhead{RA} & \colhead{Dec} & \colhead{b$^{\circ}$} & \colhead{$\Delta$mag} & \colhead{p(mas)} & \colhead{Out} & \colhead{Days} &  \colhead{SDSS} & \colhead{CRTS} & \colhead{Spec} & \colhead{Surveys} }
\startdata
 18abhypsw &  00 52 09.61 &   +43 56 20.1 & -18.9 &  17.0-20.0 &  $0.50 \pm 0.07$ &         2 &    49 &    Y &   Y2 &        --- &   ROSAT,G \\
 18acyqzew &  01 43 54.23 &   +29 01 03.8 & -32.5 &  15.8-19.0 &              --- &         1 &    42 &    Y &   Y2 &        --- &     ---   \\
 18abppgdj &  03 28 15.91 &   +61 25 17.2 &   4.0 &  14.2-20.5 &  $4.09 \pm 0.15$ &         1 &   219 &  --- &  --- &        --- &  AT2019cbz,G \\
 17aaaaqee &  04 43 17.84 &   +54 02 27.6 &   5.3 &  16.7-21.0 &  $0.25 \pm 0.04$ &         1 &    14 &  --- &  --- &        --- &          --- \\
 18acemvzp &  04 57 51.59 &   -06 10 38.2 & -28.0 &  16.4-20.4 &  $1.10 \pm 0.05$ &         1 &    14 &  --- &  --- &        --- &  ROSAT,G \\
 17aaabswj &  05 25 03.73 &   +39 33 59.1 &   2.2 &  17.6-19.6 &  $1.69 \pm 0.56$ &         2 &   193 &  --- &  --- &        --- &  G\\
 18adaifhl &  05 35 45.89 &   -08 47 48.8 & -20.8 &  18.6-20.0 &              --- &         1 &    30 &  --- &  --- &        --- &   AT2018lu,ROSAT \\
 18acxhxny &  05 47 04.15 &   +26 45 04.0 &  -0.9 &  17.2-21.0 &              --- &         1 &    29 &  --- &  --- &        --- &  --- \\
 18acnatft &  06 29 04.48 &   +11 25 37.5 &   0.3 &  17.8-21.0 &              --- &         1 &    14 &  --- &  --- &        --- &    ---   \\
 19aakncwr &  08 10 51.92 &   -04 04 28.5 &  15.6 &  16.4-20.2 &              --- &         1 &    53 &  --- &  --- &        --- &    AT2017cjw \\
 18acrhioo &  08 40 19.19 &   -03 51 24.7 &  22.1 &  17.2-19.3 &              --- &         2 &    35 &  --- &  --- &        --- &    --- \\
 19aajywdx &  09 24 34.29 &     +08 40 31 &  37.9 &  16.2-20.7 &              --- &         1 &    24 &    Y &  --- &        --- &     ---\\
 19aarflna &  11 34 02.27 &   +14 01 29.3 &  67.7 &  13.6-20.0 &              --- &         1 &    26 &    Y &  --- &        --- &    ---\\
 19aaqxmmw &  14 58 58.58 &   -06 07 05.8 &  44.7 &  15.4-20.3 &              --- &         1 &    23 &  --- &  --- &        --- &    --- \\
 18abetcrn &  17 28 03.39 &   +25 50 50.3 &  28.9 &  19.4-21.1 &              --- &         2 &   296 &  --- &  --- &        P,K &  --- \\
 18abltirl &  18 06 56.21 &   +06 10 34.8 &  12.7 &  15.9-21.0 &  $1.93 \pm 0.19$ &         2 &   341 &  --- &  --- &        --- & Swift,G \\
 18aaniudh &  18 09 47.90 &   +30 23 05.7 &  21.7 &  16,9-19.5 &  $0.84 \pm 0.03$ &         3 &   379 &  --- &   Y2 &        --- &      --- \\
 18ablmpfr &  18 44 34.47 &   +13 07 17.7 &   7.4 &  19.0-21.0 &              --- &         1 &    30 &  --- &  --- &        --- &  G\\
 18abdjqmg &  19 38 44.34 &   +35 40 31.2 &   6.8 &  17.5-19.9 &              --- &         1 &    30 &    Y &  --- &        --- &    --- \\
 19aaaaazu &  20 27 03.59 &   +19 23 07.9 & -10.9 &  15.4-19.4 &              --- &      SOB? &    35 &  --- &  --- &        --- &   --- \\
 18accpmqr &  20 55 08.85 &   -06 29 11.4 & -30.4 &  17.0-19.6 &              --- &         2 &    48 &    Y &  --- &        --- &    --- \\
\enddata
\end{deluxetable}
\end{longrotatetable}
 
\clearpage
\begin{figure}
    \centering
    \includegraphics[width=7.5in]{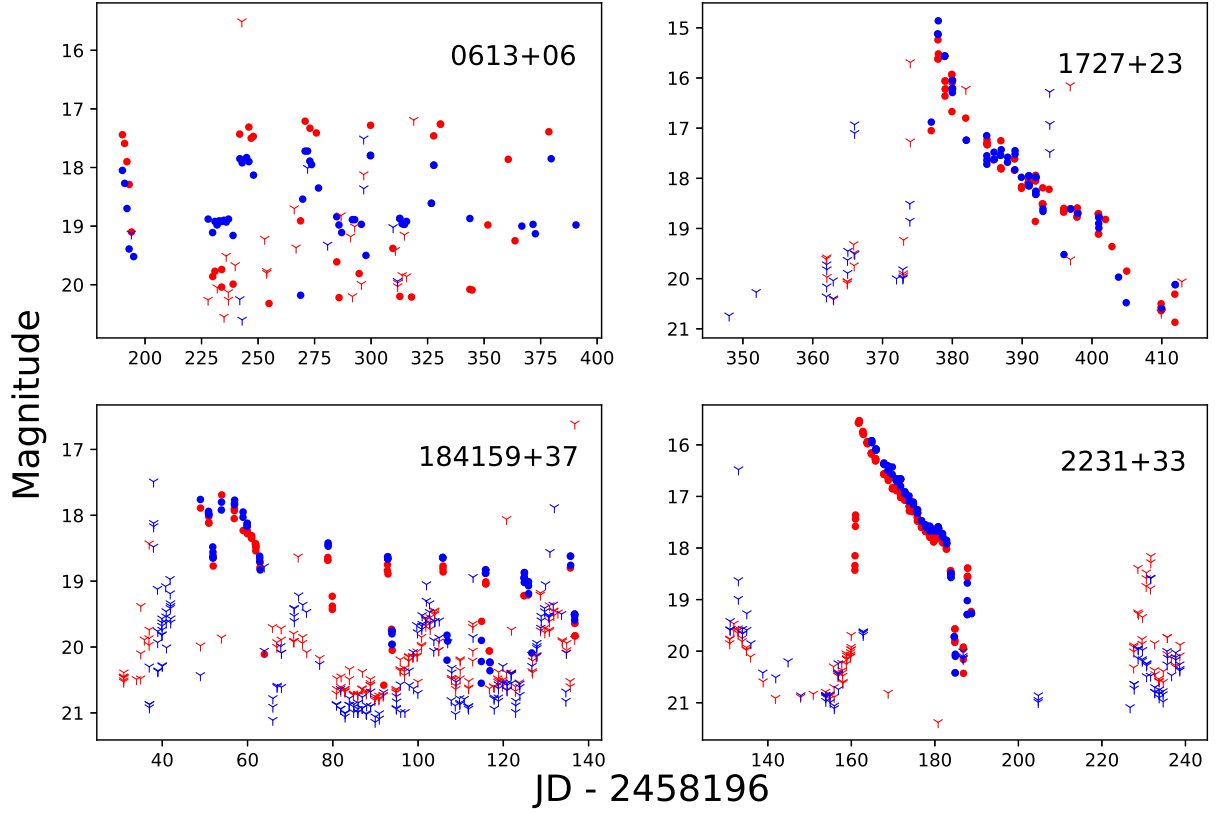}
    \caption{Examples of ZTF light curves of good CV candidates from Table 2. Filled blue and red circles are magnitudes from $g$ and $r$ filters, while light symbols are upper limits on those nights. ZTF1727+23 nad ZTF2231+33 are examples of SOBs with an outburst lasting more than 20 days and a distinctive shape. They are likely WZ Sge type systems. ZTF184159+37 shows multiple
    rebrightenings after its SOB and is a candidate for an ER UMa type system.
    ZTF0613+06 shows repeated normal outbursts with low amplitude and is likely
    a typical dwarf nova with a relatively high mass transfer rate.}
\end{figure} 

\clearpage
\begin{figure}
    \centering
    \includegraphics{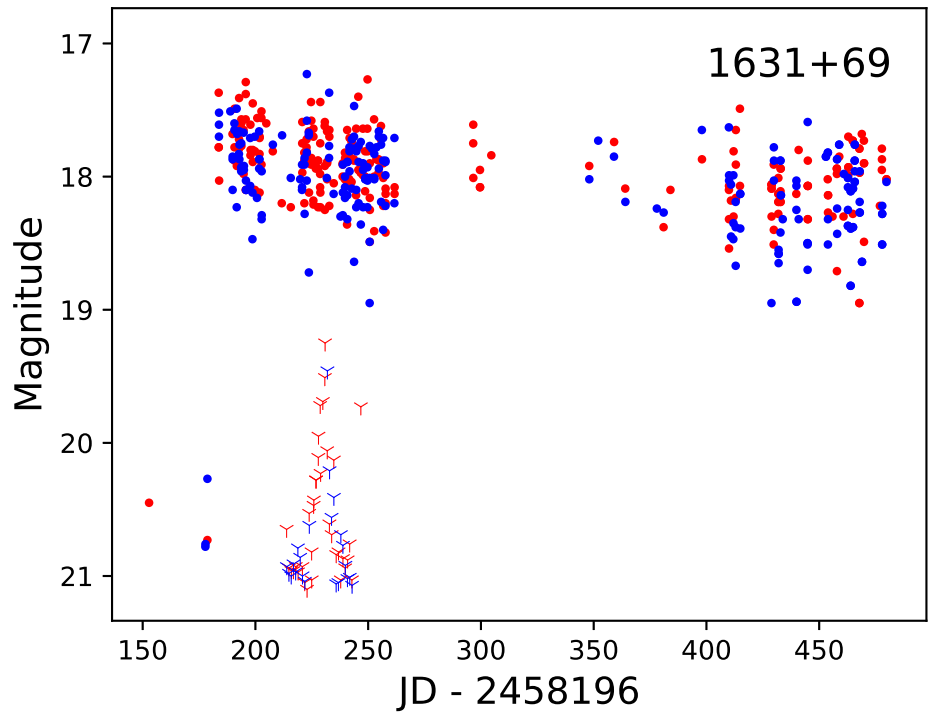}
    \caption{An example of a system showing high and low states. Symbols same
    as Figure 1.}
\end{figure}

\clearpage
\begin{figure}
  \includegraphics[height=8in]{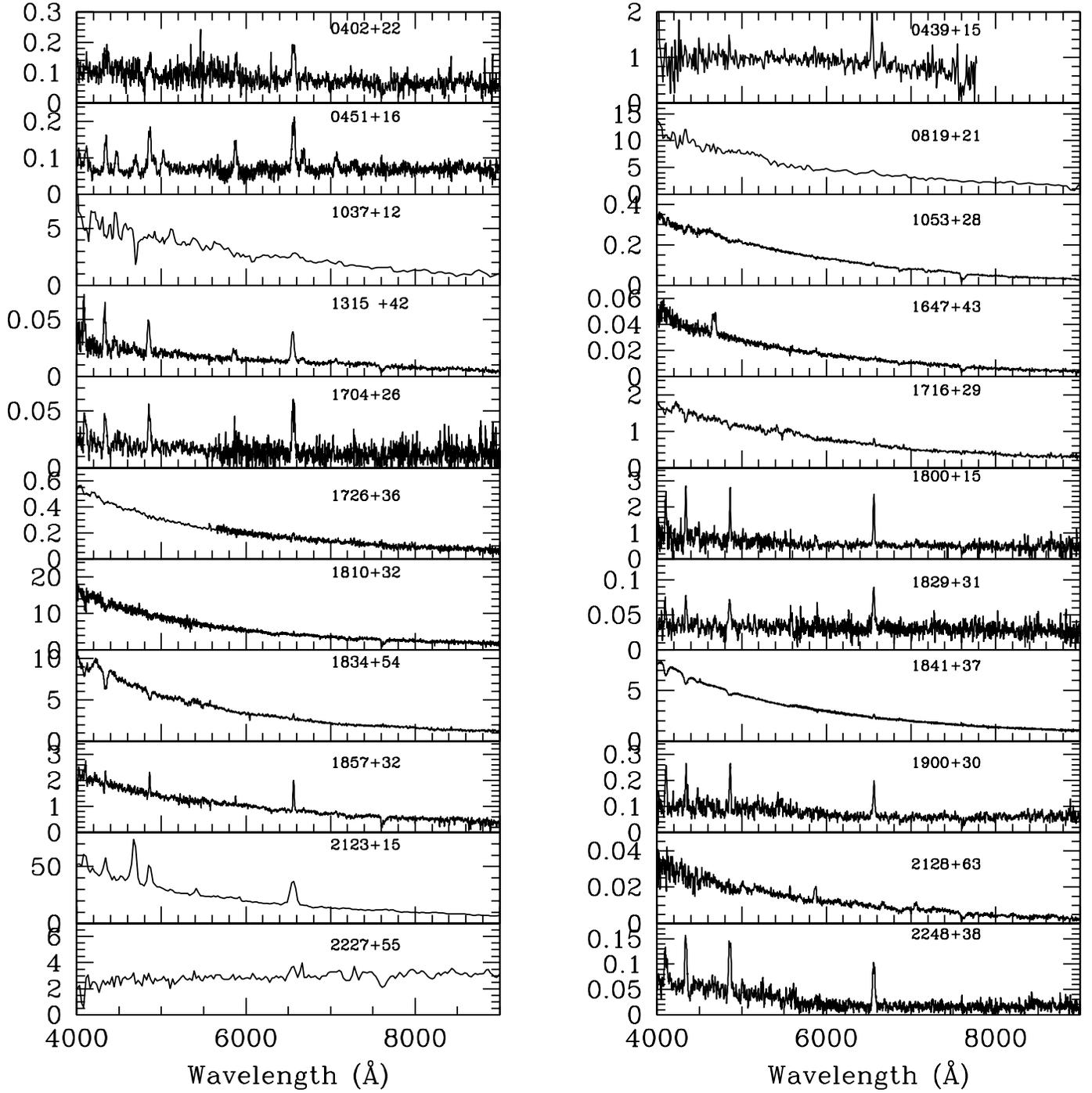}
  \caption{Blue and red spectra from Keck, Pal200in, WHT, APO, SEDM, SPRAT
  showing at least one Balmer or Helium emission line. The vertical axis is F$_{\lambda}$ in units
  of 10$^{-16}$ ergs cm${-2}$ s$^{-1}$ \AA$^{-1}$
  except for the SPRAT spectrum of ZTF0439+15 which is normalized in its reduction procedure.}
  \end{figure}
  
\clearpage
\begin{figure}
      \centering
      \includegraphics[height=6in]{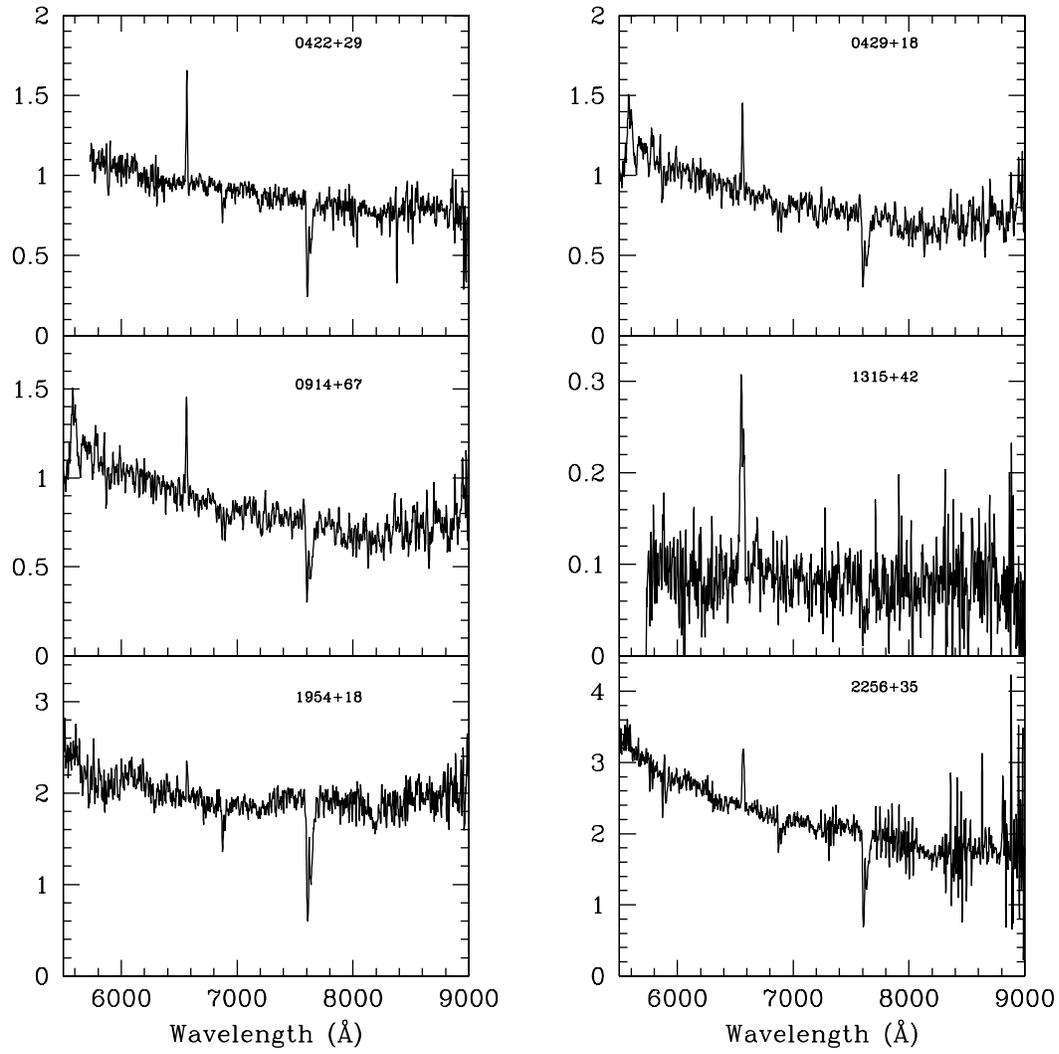}
      \caption{Red only spectra obtained at APO.The vertical axis is F$_{\lambda}$ in units
  of 10$^{-16}$ ergs cm${-2}$ s$^{-1}$ \AA$^{-1}$.}
 \end{figure}

\clearpage
\begin{figure}
    \centering
    \includegraphics{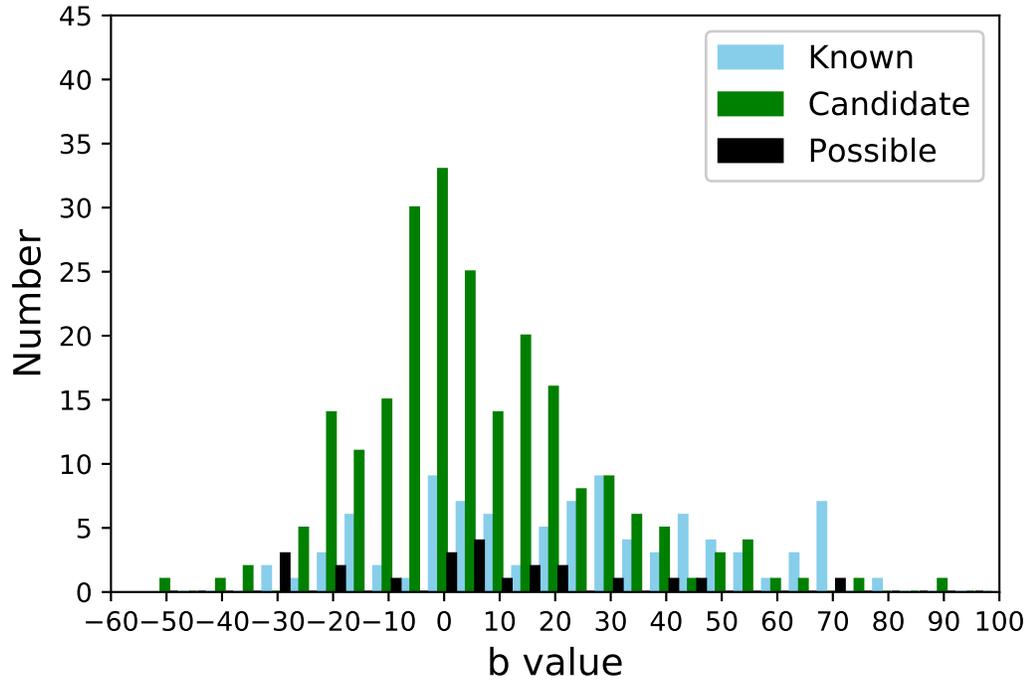}
    \caption{The number of systems in Tables 1-3 as a function of galactic latitude (in 10 deg bins).}
\end{figure}

\clearpage
\begin{figure}
    \centering
    \includegraphics{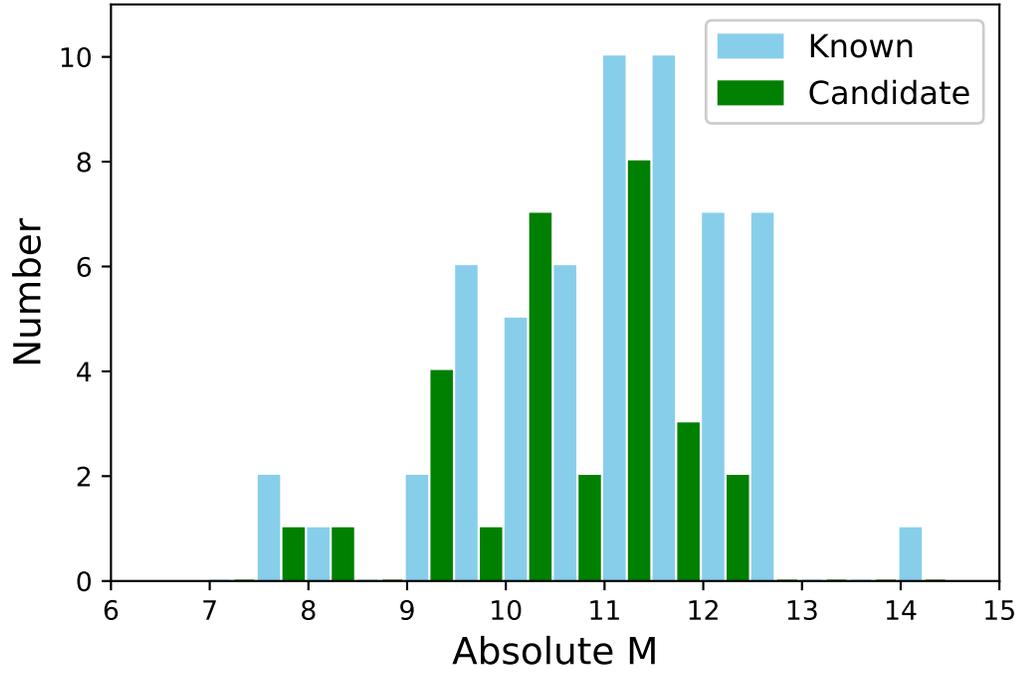}
    \caption{The number of systems in Tables 1 and 2 as a function of their absolute magnitude (in 0.5 mag bins) as determined from available Gaia parallaxes. Note that 5$\sigma$ upper limits on the magnitudes for the fainter sources means that they are only brighter limits to the true absolute magnitude at quiescence.}
\end{figure}

\clearpage
\begin{figure}
    \centering
    \includegraphics[width=6in]{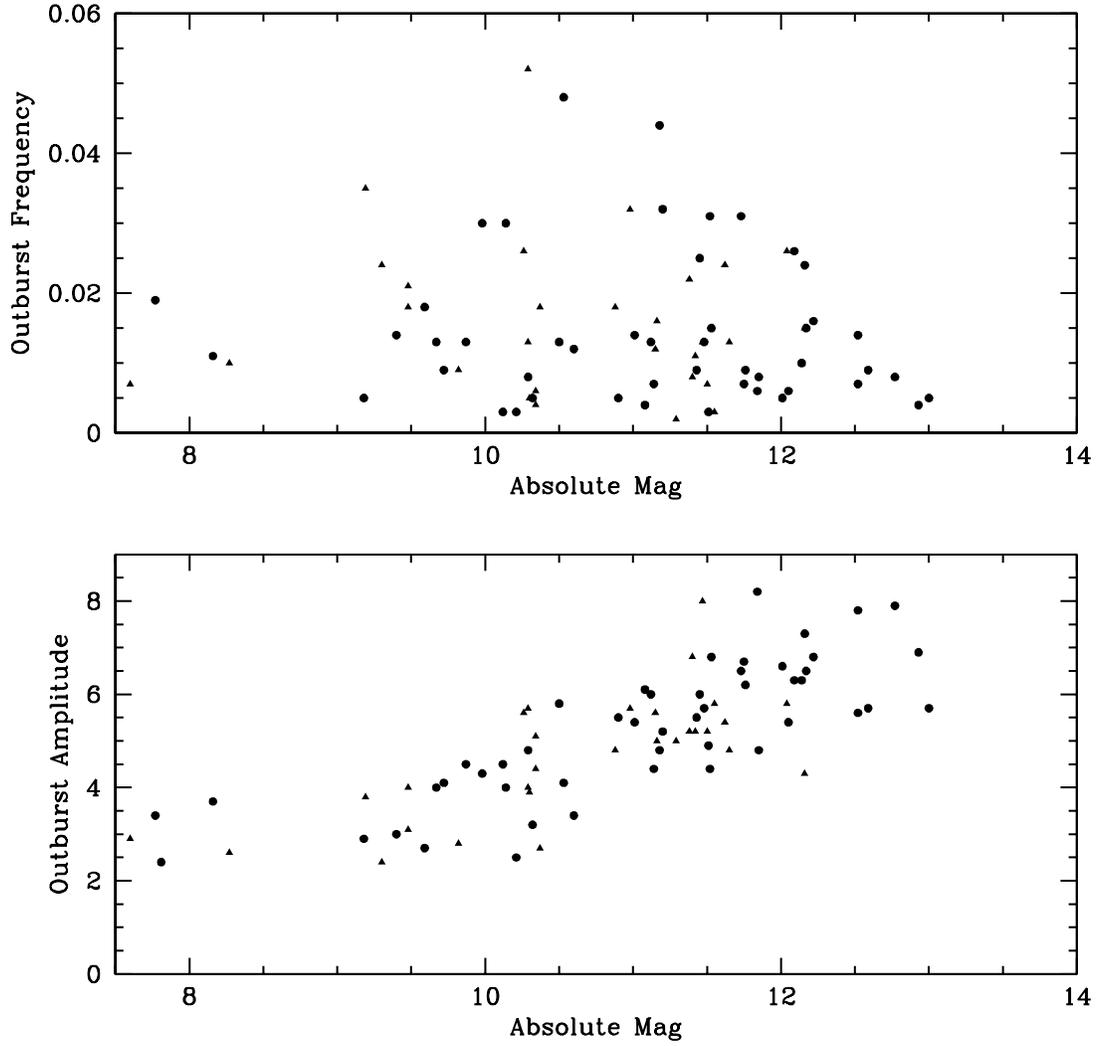}
    \caption{Plots of Absolute magnitude versus Outburst Frequency (top)
and Absolute Magnitude versus Outburst Amplitude (bottom) for the objects
showing dwarf nova type outbursts in Table 1 (solid dots) and Table 2 
(triangles).}
\end{figure}

\clearpage
\begin{figure}
    \centering
    \includegraphics[width=5in,angle=270]{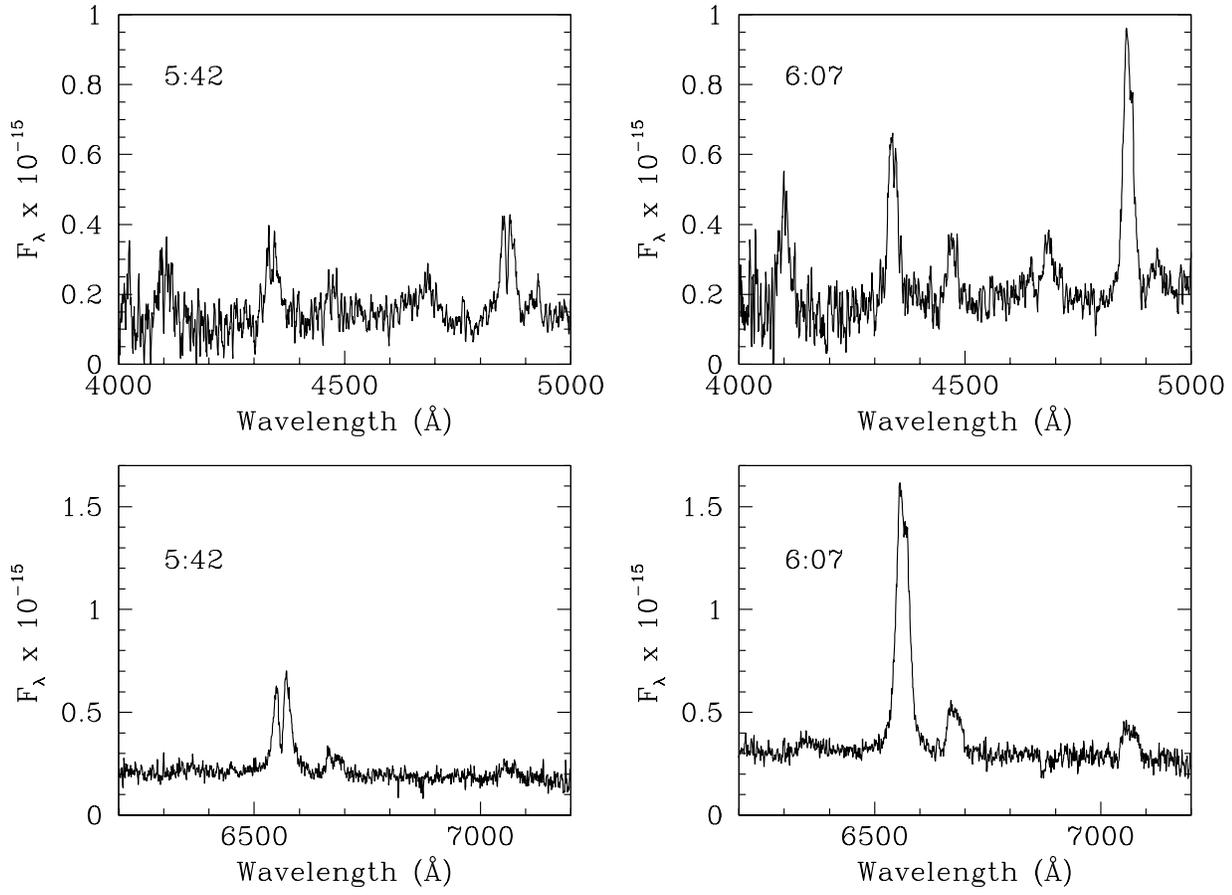}
    \caption{Two of the blue and red APO DIS spectra of 1631+69 obtained 25 min apart showing the large changes in the Balmer and He lines.}
\end{figure}

\clearpage\begin{figure}
    \centering
    \includegraphics{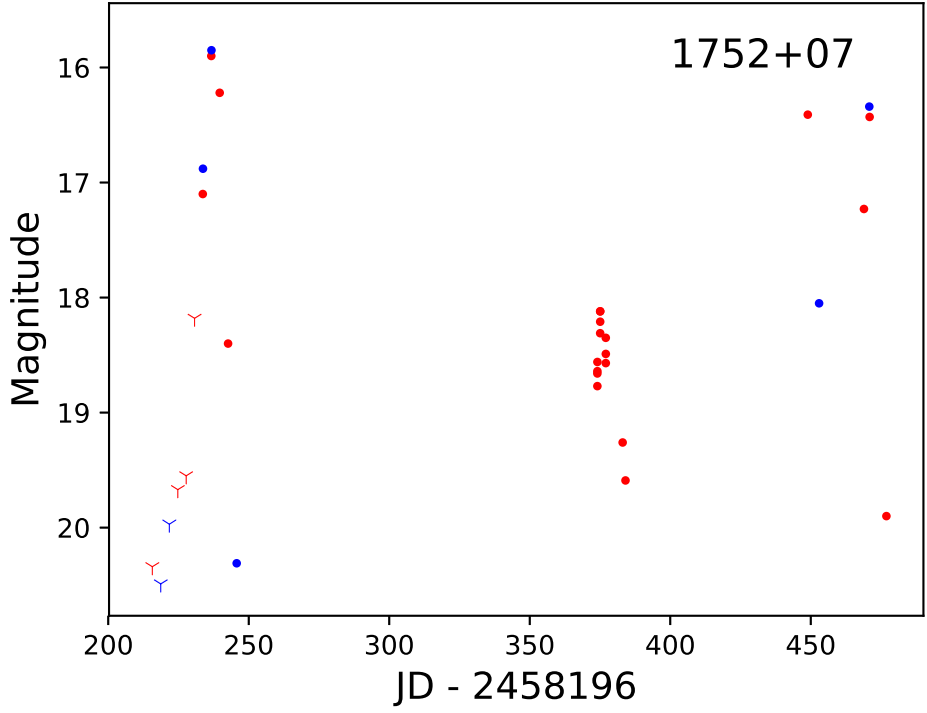}
    \caption{ZTF1752+07 (V982 Oph) light curve showing 4 dwarf nova type outbursts. Symbols same as Figure 1.}
\end{figure}


\begin{thebibliography}{}

\bibitem[Antipin \& Samus(2002)]{AS02}
Antipin, S. V. \& Samus, N. N. 2002, IBVS 5302

\bibitem[Appenzeller et al.(1998)]{A98}
Appenzeller, I., Thiering, I., Zickgraf, F. J. et al. 1998, \apjs, 117, 319 

\bibitem[Bellm et al.(2019a)]{B19a}
Bellm,E. C., Kulkarni, S. R., Graham, M. J. et al. 2019, \pasp, 131, 018002

\bibitem[Bellm et al.(2019b)]{B19b}
Bellm, E. C., Kulkarni, S. R., Barlow, T. et al. 2019, \pasp, 131, 068003

\bibitem[Benn et al. (2008)]{B08}
Benn, C., Dee, K., Agocs, T. 2008, Proc. SPIE, 7014, 70146X

\bibitem[Blagoradnova et al.(2018)]{Bl18}
Blagoradnova, N., Walters, et al. 2018, \pasp, 130, 035003

\bibitem[Breedt et al.(2014)]{Br14}
Breedt, E., G\"ansicke, B. T., Drake, A. J. et al. 2014, \mnras, 443, 3174

\bibitem[Drake et al.(2009)]{D09}
Drake, A. J., Djorgovski, S. G., Mahabel, A. et al. 2009, \apj, 696, 870

\bibitem[Drake et al.(2014)]{D14}
Drake, A. J., G\"ansicke, B. T., Djorgovski, S. G. et al. 2014, \mnras, 441, 1186

\bibitem[Gaia Collaboration (2018)]{G18}
Gaia Collaboration, 2018

\bibitem[Graham et al.(2019)]{G19}
Graham, M. J., Kulkarni, S. R., Bellm, E. C. et al. 2019, \pasp, 078001

\bibitem[Hoard et al.(1998)]{H98}
Hoard, D. W., Szkody, P., Still, M. D., Smith, R. C., Buckley, D. A. H. 1998, \mnras, 294, 689

\bibitem[Howell et al.(2001)]{HNR01}
Howell, S. B., Nelson, L. A. \& Rappaport, S. 2001, \apj, 550, 897

\bibitem[Howell et al.(1997)]{HRP97}
Howell, S. B., Rappaport, S. \& Politano, M. 1997, \mnras, 287, 929

\bibitem[Howell et al.(1995)]{HSC95}
Howell, S. B., Szkody, P., \& Cannizzo, J. K. 1995, \apj, 439, 337

\bibitem[Kasliwal et al.(2019)]{K19}
Kasliwal, M. M., Cannella, C., Bagdasaryan, A. et al. 2019, \pasp, 131, 038003

\bibitem[Lipunov et al.(2010)]{L10}
Lipunov, V., Kornilov, V., Gorbovskoy, E. et al. 2010, AdAst, 349171

\bibitem[Masci et al.(2019)]{M19}
Masci, F. J., Laher, R. R., Rusholme, B. et al. 2019, \pasp, 131, 018003

\bibitem[Mr\'oz et al.(2015)]{M15}
Mroz, P., Udalski, A., Poleski, R. et al. 2015, AcA, 65, 313

\bibitem[Oke \& Gunn (1982)]{OG82}
Oke, J. B. \& Gunn, J. E. 1982, \pasp, 94, 586

\bibitem[Oke et al. (1995)]{O95}
Oke, J. B. et al. 1995, \pasp, 107, 375

\bibitem[Osaki(1996)]{O96}
Osaki, Y., 1996, \pasp, 108, 39

\bibitem[Pala et al.(2019)]{P19}
Pala, A., G\"ansicke, B. T., Breedt, E. et al. 2019, \mnras, in press

\bibitem[Pojmanski(1997)]{P97}
Pojmanski, G. 1977, AcA, 47, 467

\bibitem[Rigault et al.(2019)]{R19}
Rigault, M., Neill, J. D., Blagoradnova, N. et al. 2019, \aap, 627, 115

\bibitem[Shappee et al.(2014)]{Sh14}
Shappee, B. J., Prieto, J. L., Grupe, D. et al. 2014, \apj, 788, 48

\bibitem[Steele et al. (2004)]{St04}
Steele, I. A., Smith, R. J., Rees, P. C. et al. 2004, Proc. SPIE, 5489, 679

\bibitem[Szkody et al.(2003)]{Sz03}
Szkody, P., Fraser, O., Silvestri, N. et al. 2003, \aj, 126, 1499

\bibitem[Szkody et al.(2011)]{Sz11}
Szkody, P., Anderson, S. F., Brooks,, K. et al. 2011, \aj, 142, 181

\bibitem[Thorstensen et al.(2002)]{T02}
Thorstensen, J. R., Patterson, J., Kemp, J., Vennes, S. 2002, \pasp, 114, 1108

\bibitem[Thorstensen et al.(1991)]{T91}
Thorsensen, J. R., Ringwalkd, F. A., Wade, R. A., Schmidt, G. D., Norsworthy, J. E. 1991, \aj, 102, 272

\bibitem[Warner(1987)]{W87}
Warner, B. 1987, \mnras, 227, 23

\bibitem[Warner(1995)]{W95}
Warner, B. 1995, Cataclysmic Variable Stars (New York: Cambridge Univ. Press)

\bibitem[York et al.(2000)]{Y00}
York, D.G., Adelman, J., Anderson, J. E. et al. 2000, \aj, 120, 1579

\end{thebibliography}
\end{document}